\documentclass[11pt,twocolumn]{article}

\usepackage[T1]{fontenc}
\usepackage[utf8]{inputenc}
\usepackage{lmodern}

\usepackage[a4paper,margin=0.75in]{geometry}
\usepackage{amsmath,amssymb,bm}
\usepackage{graphicx}
\usepackage{caption}
\usepackage{subcaption}
\usepackage{booktabs}
\usepackage{dblfloatfix} 

\usepackage{xcolor}
\usepackage[
  colorlinks=true,
  linkcolor=blue,
  citecolor=blue,
  urlcolor=blue
]{hyperref}

\usepackage[authoryear,round]{natbib}
 
\usepackage{etoolbox}



\title{\Large \textbf{A  Generative Reconstruction  of  Low-$\ell$  CMB B-Mode Signal using Reverse Diffusion in Deep Learning}}

\author{
\large
Anumanchi Agastya Sai Ram Likhit$^{*,[1]}$ \quad
Rajib Saha$^{\dagger,[1]}$ \\[0.35em]
\normalsize
$^{[1]}$Indian Institute of Science Education and Research, Bhopal \\[0.25em]
\small
$^{*}$\texttt{astropi.2003@gmail.com} \quad
$^{\dagger}$\texttt{rajib@iiserb.ac.in}
}

\date{} 

\begin{document}

\twocolumn[
\begin{@twocolumnfalse}
\maketitle

\vspace{-1 cm}
\begin{abstract}
\noindent
\normalsize
Detecting primordial B-mode polarization of the Cosmic Microwave Background (CMB) provides a direct probe of inflationary gravitational waves. However, the signal is extremely faint and contaminated by gravitational lensing, instrumental noise, and astrophysical foregrounds. Here we present a score-based diffusion approach, formulated using variance-exploding stochastic differential equations (VE-SDEs), to reconstruct the primordial B-mode angular power spectrum from contaminated observations. The method employs a reverse SDE guided by a score model trained exclusively on random realizations of the primordial low $\ell$ B-mode angular power spectrum corresponding to a fixed tensor-to-scalar ratio $r=0.001$. During inference, the reverse SDE iteratively drives the observed angular power spectrum toward the learned primordial manifold, effectively denoising and delensing the input. The model is tested on simulated observational spectra that include gravitational lensing, complex polarized foreground combinations, and instrumental noise characteristics representative of the proposed ECHO mission. The trained score model learns the underlying statistical distribution of the primordial B-mode field for the given $r$, which acts as a physics-guided prior that can generate new, consistent realizations of the signal. This approach provides a robust framework for primordial signal recovery in future CMB polarization missions.
\end{abstract}

\vspace{0.35cm}
\end{@twocolumnfalse}
]

\section{Introduction}
\label{sec:intro}
Inflation provides a robust theoretical description of the physics of the very early universe \citep{PhysRevD.23.347,Linde:1981mu,hawking1982development,PhysRevD.32.1899} and a well-motivated mechanism for generating the primordial initial conditions assumed by the $\Lambda$CDM cosmological model \citep{ratra1988cosmological,Peebles_2003}. During inflation, quantum fluctuations of the scalar field are stretched to cosmological scales, producing the primordial density (scalar) perturbations that later evolve into the large-scale structure observed today \citep{1993ppc..book.....P}. In addition to these scalar modes, inflation predicts a stochastic background of primordial gravitational waves arising from tensor perturbations of the spacetime metric \cite{PhysRevLett.78.2054}. These tensor modes leave a distinct imprint on the polarization of the Cosmic Microwave Background (CMB) in the form of a curl-like pattern known as the primordial $B$-mode polarization \citep{Kamionkowski_2016,Wang2017PrimordialGW}. Detecting this primordial $B$-mode signal would provide direct evidence for inflationary gravitational waves and place strong constraints on the energy scale of inflation through the tensor-to-scalar ratio, $r$.

However, several astrophysical and instrumental contributions obscure the primordial CMB $B$-mode signal, making its detection extremely challenging. Polarized Galactic foregrounds, such as synchrotron emission from relativistic electrons at low frequencies \citep{Kogut_2012} and thermal dust emission from magnetically aligned interstellar grains at high frequencies \citep{1999ApJ...524..867F}, produce complex, spatially varying signals that dominate over the primordial component at most frequencies. In addition to these foregrounds, gravitational lensing by large-scale structures converts CMB $E$-mode polarization into B-modes \citep{blanchard1987gravitational,bernardeau1996weak}, producing a lensing induced $B$-mode signal that is significantly larger than the expected primordial contribution. Instrumental noise and systematics further complicate the measurement \citep{o2007systematic}. 

Several ground-based, balloon-borne, and satellite experiments have attempted to measure the CMB B-mode polarization, setting stronger constraints on the tensor-to-scalar ratio $r$, but no definitive detection of the primordial signal to date. These experiments starting from WMAP \cite{Bennett_2013} and Planck \citep{2020} to dedicated B-mode instruments such as BICEP/Keck \citep{paoletti2022planck}, POLARBEAR \citep{Kermish_2012}, SPTpol \citep{Austermann_2012}, and ACTpol \citep{Naess_2014} have so far only established upper limits on $r$. The latest and most stringent constraint, $r \leq 0.036$ at the $95 \%$ confidence level, is provided by the Planck and BICEP/Keck Array 2018 analysis \citep{paoletti2022planck}. Alongside observational efforts, extensive methodological developments have been made to improve the reconstruction of the CMB B-mode polarization angular power spectrum and map. Classical approaches include parametric and template-fitting techniques, internal linear combination (ILC) methods \citep{Saha:2007gf,Sudevan:2017una,Yadav_2021}, and component-separation schemes tailored for polarized foregrounds \citep{Remazeilles_2018,Eriksen_2008,carones2024optimizingblindreconstructioncmb,steier2025unbiasedprimordialgravitationalwave}. More recently, the growth of large datasets and advances in machine learning have motivated the use of deep-learning models for denoising, delensing, and reconstructing CMB polarization signals from complex contaminations \citep{Caldeira:2018ojb,Chanda:2021qbf,Yan:2023oan,yan2024cmbfscnncosmicmicrowavebackground,pal2024accurateunbiasedreconstructioncmb,2020JCAP...07..017F,Adak:2025iyj,Sudevan:2024hwq}. Most of these approaches rely on supervised learning, where networks are trained on simulated foregrounds or lensing contaminants so that the model learns their characteristic structure and removes them from the observed maps. However, such methods are inherently limited by their dependence on the training data: if the true foregrounds or contaminants differ from those represented in the training set, the model’s performance can degrade significantly.

An emerging direction involves generative models, particularly diffusion-based
approaches, in astrophysics and cosmology \cite{boruah2025diffusionbasedmassmapreconstruction,zhao2023diffusionmodelconditionallygenerate,lizarraga2025understandinggalaxymorphologyevolution,Riveros_2025}, which have demonstrated strong potential for CMB reconstruction. Recent work has used diffusion models for denoising and delensing of non-Gaussian CMB lensing potential fields \citep{flöss2024denoisingdiffusiondelensingdelight}. Score-based generative models, which learn the gradient of the log-probability density (the score) of the underlying data distribution, provide an alternative framework for reconstructing the primordial $B$-mode signal. By training the model to estimate the score of the CMB $B$-mode distribution at multiple noise levels, one can iteratively sample from the learned score field to obtain a denoised and delensed estimate of the signal. In addition to reconstruction, such models can also be used to generate high-fidelity simulations consistent with the learned distribution for a given $r$. The next generation of CMB polarization experiments such as CMB-S4 \citep{abazajian2022cmb}, the Simons Observatory \citep{namikawa2022simons}, LiteBIRD \citep{litebird2023probing}, and the proposed ECHO (CMB-Bharat) \citep{Adak_2022} mission are designed to achieve sensitivities capable of probing primordial gravitational waves down to  $\sim 10^{-3}$. These experiments will provide the precision required to isolate the primordial $B$-mode signal, provided that delensing and foreground mitigation are performed with correspondingly high accuracy. In this work, we develop a score-based stochastic differential equation framework to jointly denoise and delense the CMB $B$-mode signal and reconstruct its angular power spectrum in the low-multipole regime ($\ell < 32$) using simulated observations representative of the ECHO mission.

The remainder of this paper is organized as follows. In Section 2, we describe the methodology, including the score-based stochastic differential equation framework and the process of reconstruction. Section 3 describes the simulations used to train and evaluate the method, including the generation of lensed CMB $B$-modes, complex polarized foregrounds, and realistic instrumental effects at $r=0.001$. In Section 4, we detail the model training procedure, network specifications, and we present the reconstruction results along with a series of robustness tests. Finally, Section 5 provides a summary of our findings and the main conclusions.

\section{Formalism}
We employ score-based diffusion models \citep{song2020generativemodelingestimatinggradients} formulated through a variance--exploding stochastic differential equation (VE-SDE) \citep{song2021scorebasedgenerativemodelingstochastic,DBLP:journals/corr/abs-2011-13456} to reconstruct primordial B-mode power spectra from contaminated CMB observations at \(r=0.001\). 
Score-based generative models learn the gradient of the log-probability density of the data distribution (the \emph{score} ) \(\nabla_{\mathbf{x}} \log p(\mathbf{x})\), and enable sampling via a reverse diffusion process, in which an initially noise-dominated distribution is progressively driven back toward the data manifold. In our setup, a realization of the primordial $B$-mode angular power spectrum vector \(\mathbf{x}_0\in\mathbb{R}^{30}\) is progressively perturbed with Gaussian noise over diffusion times \(t\in[0,1]\), such that the data distribution is continuously driven toward a highly corrupted, noise-dominated reference distribution at large \(t\).

\subsection{Forward and reverse SDEs}

A diffusion process in continuous time is governed by the forward stochastic differential equation
\begin{equation}
    d\mathbf{x} = f(\mathbf{x}, t)\,dt + g(\mathbf{x}, t)\,d\mathbf{w},
\end{equation}
where \(f\) is the drift coefficient, \(g\) is the diffusion coefficient, and \(d\mathbf{w}\) denotes the increment of a Wiener process. The distribution induced by the forward process at time \(t\) is denoted \(p_t(\mathbf{x})\).Given this forward SDE, \cite{Anderson1982ReversetimeDE} established that a corresponding reverse-time SDE exists:

\begin{figure*}[t]
    \centering
    \includegraphics[width=1\linewidth]{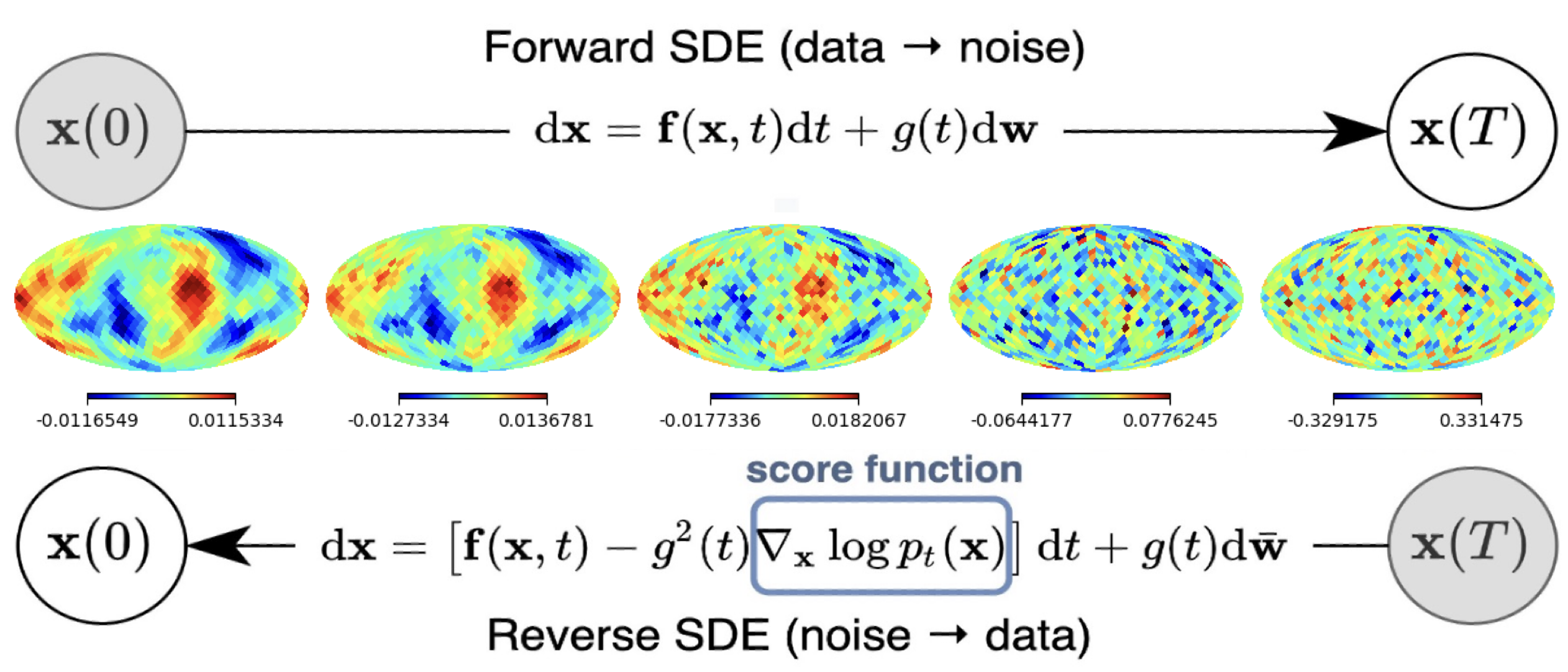}
    \caption{Top panel shows transformation of the data to a noise dominated distribution by slowly adding noise to it using a forward SDE. The bottom panel shows that samples of noise minimized data can be generated by using the reverse SDE method. In both cases the differential equations used are mentioned. The reverse SDE computation requires knowledge of score function.}
    \label{fig:sdersde}
\end{figure*}
\begin{align}
d\mathbf{x}
&=
\left[
    f(\mathbf{x}, t)
    - g(\mathbf{x}, t)^2\,
      \nabla_{\mathbf{x}} \log p_t(\mathbf{x})
\right] dt
\nonumber \\
&\quad
+ g(\mathbf{x}, t)\, d\bar{\mathbf{w}},
\label{eq:reverse_sde}
\end{align}

\noindent where \(d\bar{\mathbf{w}}\) is a Wiener process evolving backward in time. This reverse SDE transforms the noise-dominated terminal distribution back toward the data distribution as shown in Fig.\ref{fig:sdersde}, provided the score \(\nabla_{\mathbf{x}} \log p_t(\mathbf{x})\) is available. For this work, we adopt the variance–exploding SDE, commonly used in score-based generative modeling. In this formulation, the drift vanishes and the diffusion magnitude grows monotonically with time:
\begin{equation}
    f(\mathbf{x}, t) = 0, \qquad
    g(t)
    = \sigma(t)\,
      \sqrt{
        2 \ln\!\left(\frac{\sigma_{\max}}{\sigma_{\min}}\right)
      }.
\end{equation}
with an exponential noise schedule
\begin{equation}
    \sigma(t)
    = \sigma_{\min}
      \left(\frac{\sigma_{\max}}{\sigma_{\min}}\right)^{t},
    \qquad t\in[0,1].
\end{equation}

\noindent The noise scales $\sigma_{\min}$ and $\sigma_{\max}$ are selected following
standard practice in variance exploding SDEs, with $\sigma_{\min}$ introducing
negligible perturbations at early times and $\sigma_{\max}$ ensuring that the
forward process reaches a highly noise--dominated reference distribution at $t = 1$ \citep{song2021scorebasedgenerativemodelingstochastic}.
 The forward process therefore simplifies to the form
\begin{equation}
    d\mathbf{x}
    = g(t)\, d\mathbf{w}
    = \sigma(t)
      \sqrt{
        2 \ln\!\left(\frac{\sigma_{\max}}{\sigma_{\min}}\right)
      }\, d\mathbf{w},
\end{equation}
which gradually increases the variance of the perturbed sample while preserving the mean, eventually producing a highly-variance reference distribution. By substituting this diffusion coefficient into the general reverse-time formulation, the reverse SDE becomes

\begin{align}
d\mathbf{x}
&=
-2\,\sigma(t)^2
\ln\!\left(\frac{\sigma_{\max}}{\sigma_{\min}}\right)
\nabla_{\mathbf{x}}\log p_t(\mathbf{x})\, dt
\nonumber \\
&\quad
+ \sigma(t)
\sqrt{2\ln\!\left(\frac{\sigma_{\max}}{\sigma_{\min}}\right)}
\, d\bar{\mathbf{w}}.
\label{rsde}
\end{align}

\noindent These equations form the basis of the generative sampling procedure used in this work.

\subsection{Score network and denoising score matching}
To utilize the reverse SDE in Eq.~\ref{rsde}, we require the score
\(\nabla_{\mathbf{x}} \log p_t(\mathbf{x})\), which in general is not analytically traceable. We approximate it using deep learning model
\(s_\theta(\mathbf{x}_t,t)\), referred to as \texttt{ScoreNet1D}. We design a
Multilayer Perceptron (MLP)-based \citep{rosenblatt1958perceptron} score estimator that takes as input:
\begin{enumerate}
    \item A vector \(\mathbf{x}\in\mathbb{R}^{30}\) representing the
    \(B\)-mode power spectrum in the multipole range \(2 \leq \ell \leq 31\), and
    \item A scalar diffusion time \(t\), embedded using random Fourier
    features following \citep{song2021scorebasedgenerativemodelingstochastic}
    to encode the noise scale of the variance–increasing SDE.
\end{enumerate}

\noindent The network is trained to approximate the marginal score
\(\nabla_{\mathbf{x}_t}\log p_t(\mathbf{x}_t)\) at each diffusion time
\(t\). We employ a variance-exploding SDE, which induces a Gaussian
perturbation of the form
\begin{equation}
    \mathbf{x}_t
    =
    \mathbf{x}_0 + \sigma_{\mathrm{marg}}(t)\,\mathbf{z},
    \quad
    \mathbf{z}\sim\mathcal{N}(\mathbf{0},\mathbf{I}),
\end{equation}
so that the conditional distribution is
\(p(\mathbf{x}_t|\mathbf{x}_0)
  = \mathcal{N}(\mathbf{x}_0, \sigma_{\mathrm{marg}}(t)^2\mathbf{I})\).
The corresponding conditional score takes the form
\begin{equation}
\label{eq21}
    \nabla_{\mathbf{x}_t}\log p(\mathbf{x}_t|\mathbf{x}_0)
    =
    -\frac{\mathbf{x}_t - \mathbf{x}_0}{\sigma_{\mathrm{marg}}(t)^2}
    =
    -\frac{\mathbf{z}}{\sigma_{\mathrm{marg}}(t)}.
\end{equation}

\noindent During training, the score network is optimized using a weighted
denoising score-matching (DSM) objective \citep{6795935}. The general weighted
DSM loss takes the form
\begin{align}
\mathcal{L}_{\mathrm{WDSM}}(\theta)
&=
\mathbb{E}_{t,\mathbf{x}_0,\mathbf{z}}
\nonumber \\
&\quad \times
\Bigl[
\lambda(t)\,
\bigl\|
s_\theta(\mathbf{x}_t,t)
-
\nabla_{\mathbf{x}_t}
\log p(\mathbf{x}_t|\mathbf{x}_0)
\bigr\|^2
\Bigr].
\label{eq:wdsm_general}
\end{align}

\noindent which, after substituting Eq \ref{eq21},
becomes
\begin{align}
\mathcal{L}_{\mathrm{WDSM}}(\theta)
&=
\mathbb{E}_{t,\mathbf{x}_0,\mathbf{z}}
\nonumber \\
&\quad \times
\Bigl[
\lambda(t)\,
\bigl\|
s_\theta(\mathbf{x}_t,t)
+
\frac{\mathbf{z}}{\sigma_{\mathrm{marg}}(t)}
\bigr\|^2
\Bigr].
\label{eq:wdsm_gaussian}
\end{align}

\noindent Following \citep{song2021scorebasedgenerativemodelingstochastic}, we set
\(\lambda(t) = \sigma_{\mathrm{marg}}(t)^2\). This choice balances the
contribution of different noise levels and yields
\begin{equation}
\label{eq:dsm_final}
\mathcal{L}(\theta)
=
\mathbb{E}_{t,\mathbf{x}_0,\mathbf{z}}
\left[
\left\|
\sigma_{\mathrm{marg}}(t)\,
s_\theta(\mathbf{x}_t,t)
+
\mathbf{z}
\right\|^2
\right].
\end{equation}

\noindent Minimization of Eq.~\eqref{eq:dsm_final} ensures that the learned function
\(s_\theta(\mathbf{x}_t,t)\) converges to the marginal score
\(\nabla_{\mathbf{x}_t}\log p_t(\mathbf{x}_t)\) for all diffusion times \(t\).

\subsection{Euler--Maruyama discretization}

To numerically simulate both the forward and reverse variance–increasing SDEs, we discretize the time interval \([0,1]\) into \(N\) uniform steps of size \(\Delta t = 1/N\). The discrete time points are \(t_k = k\,\Delta t\) for \(k = 0,1,\dots,N\), where \(t_N = 1\) is the terminal time. At each step, the Brownian increment is approximated as
\begin{equation}
\Delta \mathbf{w}_k = \ \mathbf{w}_{t_{n+1}}  -  \mathbf{w}_{t_{n}} \sim\sqrt{\Delta t}\,\xi_k, \quad \xi_k \sim \mathcal{N}(0,1).
\end{equation}

\noindent For the forward variance–increasing SDE
\[
d\mathbf{x} = g(t)\, d\mathbf{w},
\qquad
g(t) = \sigma(t)
\sqrt{
2\ln\!\left(\frac{\sigma_{\max}}{\sigma_{\min}}\right)
},
\]
the Euler--Maruyama update becomes
\begin{align}
\mathbf{x}_{k+1}
&=
\mathbf{x}_k
+
g(t_k)\,\Delta\mathbf{w}_k
\nonumber \\
&=
\mathbf{x}_k
+
\sigma(t_k)
\sqrt{
2\ln\!\left(\frac{\sigma_{\max}}{\sigma_{\min}}\right)
}\,
\sqrt{\Delta t}\,
\boldsymbol{\xi}_k.
\label{eq:em_forward}
\end{align}

\noindent with the instantaneous noise schedule
\[
\sigma(t_k)
=
\sigma_{\min}
\left(\frac{\sigma_{\max}}{\sigma_{\min}}\right)^{t_k}.
\]

\noindent For the reverse-time SDE associated with the variance–increasing process,
\begin{align*}
d\mathbf{x}
&=
-2\,\sigma(t)^2\,
\ln\!\left(\frac{\sigma_{\max}}{\sigma_{\min}}\right)
\nabla_{\mathbf{x}}\log p_t(\mathbf{x})\, dt
\nonumber \\
&\quad
+ \sigma(t)
\sqrt{
2\ln\!\left(\frac{\sigma_{\max}}{\sigma_{\min}}\right)
}
\, d\bar{\mathbf{w}}.
\end{align*}
the Euler--Maruyama discretization yields
\begin{align}
\label{eq:em_reverse}
\mathbf{x}_{k+1}
=
&\;
\mathbf{x}_k
-
2\,\sigma(t_k)^2\,
\ln\!\left(\frac{\sigma_{\max}}{\sigma_{\min}}\right)
\nabla_{\mathbf{x}}\log p_{t_k}(\mathbf{x}_k)\,\Delta t \nonumber\\
&\quad+
\sigma(t_k)
\sqrt{
2\ln\!\left(\frac{\sigma_{\max}}{\sigma_{\min}}\right)
}\,
\sqrt{\Delta t}\,\boldsymbol{\xi}_k.
\end{align}

\noindent These discrete updates provide practical numerical approximations to the continuous forward and reverse SDEs, enabling step-by-step generation of samples from the learned score model.

\section{Simulations}
\label{sec:simulations}

To train and evaluate the score-based reconstruction model, we simulate both
(1) pure primordial CMB $B$-mode angular power spectrum for learning the score prior and
(2) realistic observational spectra that include gravitational lensing,
instrumental noise, and polarized Galactic foregrounds representative of the
Exploring Cosmic History and Origin (ECHO) mission\footnote{\href{https://cmb-bharat.in/}{https://cmb-bharat.in/}} \citep{Adak_2022}. In this section, we first describe the generation of the noise-free primordial training dataset, followed by beam convolution consistent with the ECHO instrument specifications. We then present the construction of the testing dataset, which combines primordial $B$-modes with lensing-induced $B$-modes, ECHO-like instrumental noise, and complex polarized foreground emission models.

\subsection{Training Data: Primordial CMB $B$-mode Realizations}

We generate the theoretical primordial $B$-mode angular power spectrum
$C_\ell^{BB(\mathrm{tensor})}$ using the CAMB Boltzmann code
\citep{Lewis_2000}, adopting the best-fit $\Lambda$CDM parameters from the
Planck 2018 data release \cite{2020} with a fixed tensor-to-scalar ratio
$r=0.001$. HEALPix ( Hierarchical Equal Area iso-Latitude Pixelization) \citep{Gorski_2005} is then used to simulate full-sky
Stokes-$Q/U$ maps from these spectra for a large number of independent random
seeds. The corresponding power spectra are extracted using spherical harmonic
transforms to obtain $50{,}000$ independent realizations of the primordial
multipole range $2 \le \ell \le 31$. To incorporate the angular resolution of the instrument, each realization is
convolved with a symmetric Gaussian beam characterized by the transfer
function
\begin{equation}
    b_{\ell}
    = \exp\!\left[-\frac{\ell(\ell+1)\sigma_b^2}{2}\right].
\end{equation}
The Gaussian width $\sigma_b$ is defined in terms of the full width at half
maximum (FWHM) by
\begin{equation}
    \sigma_b
    = \frac{\theta_{\mathrm{FWHM}}}{\sqrt{8\ln 2}},
\end{equation}
where we adopt the 28\,GHz channel specification of the proposed
ECHO mission~\citep{Adak_2022}, corresponding to
\(\theta_{\mathrm{FWHM}} = 24.8'\).
The beam-convolved multipoles \(b_\ell^2 C_\ell^{BB}\) constitute the complete
training dataset used to learn the marginal score of the primordial
distribution. Crucially, these spectra include no lensing, instrumental noise,
or foreground contamination, ensuring that the model learns an unbiased prior
for the primordial $B$-mode statistics.

\subsection{Testing Data: Realistic Observational Spectra}
\label{subsec:testing-data}

To evaluate reconstruction performance under realistic conditions, we generate a
separate testing dataset of $\sim 20{,}000$ CMB $B$-mode realizations that do
not share seeds with the training set. Each realization consists of a sum of
primordial tensor modes and lensing-induced modes,
\begin{equation}
C_\ell^{BB(\mathrm{total})}
=
C_\ell^{BB(\mathrm{tensor})}
+
C_\ell^{BB(\mathrm{lensing})},
\end{equation}
where
$C_\ell^{BB(\mathrm{lensing})}$ is computed with CAMB by setting the lensing
amplitude $A_{\mathrm{lens}}=1$. The pure primordial component serves as the
ground truth for validation.

\subsubsection{Instrumental noise}
To simulate instrument-induced uncertainties, we adopt the expected performance
of the ECHO mission at its low–frequency CMB polarization channel. In
particular, we use:
\begin{align*}
\nu &= 28~\mathrm{GHz}, \nonumber \\
\theta_{\mathrm{FWHM}} &= 24.8', \nonumber \\
\sigma_{\mathrm{rms}} &= 16.5~\mu\mathrm{K}\!\cdot\mathrm{arcmin}.
\end{align*}

\noindent The pixel-space noise variance is computed as
\begin{equation}
\sigma_v^2
=
\left(
\frac{\sigma_{\mathrm{rms}}\;\pi}{60\times180}
\right)^2
\left(\frac{N_{\mathrm{pix}}}{4\pi}\right),
\end{equation}
and Gaussian-distributed $Q/U$ noise maps are generated as
\[
Q_{\mathrm{noise}}(\hat{n}),\;
U_{\mathrm{noise}}(\hat{n})
\sim \mathcal{N}(0,\sigma_v^2).
\]
The noise multipoles $N_\ell^{BB}$ are obtained via harmonic transformation and
added to the simulated spectra to reflect realistic detector sensitivity.

\subsubsection{Polarized foreground contamination } 
We model polarized Galactic foreground emission at 28\,GHz using the
\texttt{PySM3} sky simulation framework \citep{Thorne_2017,Zonca_2021,Panexp_2025,yan2025foregroundremovalgroundbasedcmb}, which provides
physically motivated, data-constrained models of diffuse microwave emission.
PySM has been extensively used in CMB foreground studies and forecasting
analyses for current and future polarization experiments \citep{2025,Joseph_2023,m2025deeplearningcmbforeground}. At low multipoles, the dominant polarized foregrounds are Galactic synchrotron
radiation and thermal dust emission, with a subdominant contribution from
anomalous microwave emission (AME). Synchrotron emission arises from relativistic
electrons spiraling in the Galactic magnetic field and dominates at low
frequencies \citep{smoot1999synchrotronradiationcmbforeground,Weiland_2022}, while thermal dust emission originates from magnetically aligned
interstellar grains and becomes increasingly important at higher frequencies
\citep{Prunet:1999mh}. Although AME is weakly polarized, it can still
contribute non-negligible contamination on large angular scales
\citep{Ysard_2010,Dobler_2008}.

We define four foreground configurations (F1--F4), representing different
combinations of PySM polarized foreground models, designed to span a
representative range of spectral and spatial complexity:
\[
\begin{aligned}
\mathrm{F1}&: (a2,\,s1,\,d1), \qquad
\mathrm{F2}: (a2,\,s1,\,d4),\\
\mathrm{F3}&: (a2,\,s3,\,d1), \qquad
\mathrm{F4}: (a2,\,s3,\,d4),
\end{aligned}
\]
where $s1$ and $s3$ correspond to alternative synchrotron models with increasing
spatial and spectral complexity, $d1$ and $d4$ represent thermal dust models with
different prescriptions for spatial variation of the dust spectral index, and
$a2$ denotes a polarized AME component \citep{2008AA4901093M,2015MNRAS.451.4311R,2013AA553A96D,2012ApJ...753..110K}. For a more detailed description of these foreground model families and their parameterizations, we refer the reader to the PySM documentation\footnote{\href{https://pysm3.readthedocs.io/en/latest/models.html}{https://pysm3.readthedocs.io/en/latest/models.html}}.

For each foreground configuration, we generate $5{,}000$ foreground-contaminated $B$-mode spectra, yielding a total of $20{,}000$ mock foreground realizations across F1--F4. These foreground spectra enter the observed signal exclusively through the additive foreground power term $F_\ell^{BB}$ in
Eq.~\eqref{eq:obs_model}, and are subsequently combined with primordial tensor modes, lensing-induced $B$-modes, and instrumental noise to construct the final test dataset. Importantly, no foreground information is provided to the score-based model during training or inference; the foreground contribution is treated entirely as an unmodeled contaminant. This setup enables a direct and
conservative assessment of whether the learned primordial score prior alone is
sufficient to guide the reverse VE--SDE toward accurate recovery of the low-$\ell$ primordial $B$-mode signal under realistic observational conditions.

\subsubsection{Final observed test spectra}
The observed CMB B mode power spectrum consists of various components apart from the primordial tensor mode. The complete angular  spectrum can be modelled as

\begin{align}
\hat{C}_\ell^{BB(\mathrm{obs})}
&=
\Bigl(
\hat{C}_\ell^{BB(\mathrm{tensor})}
+
C_\ell^{BB(\mathrm{lensing})}
\nonumber \\
&\quad
+
F_\ell^{BB}
+
\hat{N}_\ell^{BB}
\Bigr)
\, B_\ell^2,
\label{eq:obs_model}
\end{align}

\noindent where $F_\ell^{BB}$ and $\hat{N}_\ell^{BB}$ denote the angular power spectra of
foreground emission and instrumental noise, respectively, and $B_\ell$
represents a Gaussian beam window function of FWHM $9^\circ$. Such a choice of FWHM makes sure that we have about three pixels at $\mathrm{nside}=16$ inside each FWHM angular separation on the sky, which satisfies condition of sampling rate beyond the Nyquist frequency. In practice, for a high resolution map as  will be observed by the next generation satellite missions, the noise term is not smoothed by the native beam window function of each frequency. Only the sky  signals are smoothed by the instrumental response function. However, in this work we have chosen to work on large angular scale analysis of the B mode with gaussian smoothing of FWHM = 9 degree. The noise term should then be smoothed by the effective window function which is the ratio of 9 degree and original $24.8'$ $b_{\ell}$'s  to take into account beam deconvolution from the orginal instrumental resolution and convolution again by the 9 degree Gaussian window  function. However at low $\ell$ ($\ell \leq 31$) this ratio is identical to the $9^\circ$ beam window $B_\ell$'s for all practical purposes. The smoothing of all other signal components remains $9^\circ$ due to initial deconvolution  and final convolution process. This dataset is constructed using a general and physically motivated observational model, which provides an essential and robust foundation for evaluating whether the learned score model can satisfactorily recover the weak primordial $B$-mode signal.

\section{Model Training and Validation}
\label{sec:train-validate}

In this section, we describe the training configuration of the proposed
\texttt{ScoreNet1D} model and present a comprehensive evaluation of its
reconstruction performance. Subsection~\ref{subsec:training} details the
architecture, hyperparameters, and optimization strategy used to learn the
underlying score function of the primordial $B$-mode distribution. 
Subsection~\ref{subsec:validation} then introduces a series of statistical
tests designed to assess the accuracy, uncertainty calibration, and likelihood
consistency of the reconstructed angular power spectra.

\subsection{Model Training}
\label{subsec:training}

We train the \texttt{ScoreNet1D} model on $50{,}000$ independent realizations of the
\emph{pure} primordial CMB $B$-mode angular power spectrum generated at $r=0.001$,
as described in Section~\ref{sec:simulations}.
\begin{figure}[h]
    \centering
    \includegraphics[width=1\linewidth]{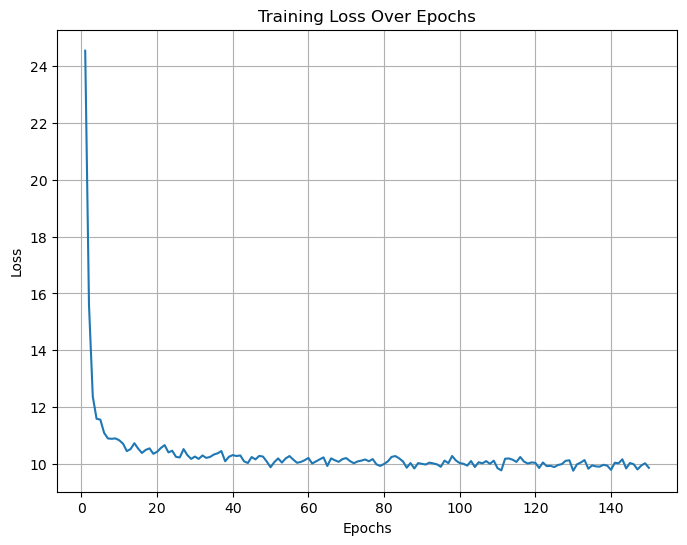}
    \caption{ Training loss evolution for \texttt{ScoreNet1D} under the weighted denoising score-matching objective (Eq.~\eqref{eq:dsm_final}) over 150 epochs. The loss decreases rapidly during the initial phase and saturates after $\sim$100 epochs, indicating stable convergence of the learned score estimator. Small fluctuations
around the plateau arise from stochastic mini-batch sampling of diffusion times and noise realizations.}
    \label{fig:loss}
\end{figure}The network is optimized using the
Adam optimizer~\cite{kingma2017adammethodstochasticoptimization} with a learning
rate of $7\times10^{-4}$ for 150 epochs and a mini-batch size of $32$. The model
takes as input a 30-dimensional vector
$\mathbf{x}\in\mathbb{R}^{30}$ corresponding to the multipole range
$2\leq \ell\leq 31$, and the diffusion time $t\in[0,1]$ embedded via random
Fourier features. The MLP architecture consists of five hidden layers of width $256$, with
Swish activation functions \citep{ramachandran2017searchingactivationfunctions} and a final linear layer producing a 30-dimensional
score estimate. The output is divided by the instantaneous noise amplitude
$\sigma_{\mathrm{marg}}(t)$ to ensure correct scaling across diffusion times,
as prescribed in the VE-SDE formulation. Training convergence is monitored using the weighted denoising score-matching
loss in Eq.~\eqref{eq:dsm_final}. Fig.\ref{fig:loss} shows the evolution of
the loss over epochs, demonstrating stable optimization and smooth convergence. This trained score model serves as the core component of our reverse VE-SDE reconstruction pipeline described next.

\subsection{Reconstruction Results and Statistical Validation}
\label{subsec:validation}
We now evaluate the performance of the proposed score-based VE-SDE framework
in reconstructing the primordial CMB $B$-mode power spectrum from contaminated
observations. We assess both the accuracy and statistical robustness of the
reconstructions through a set of complementary tests, including: (i)
visual recovery of individual realizations, (ii) ensemble-averaged power
spectra across all foreground conditions, and (iii) statistical consistency
checks involving variance calibration, likelihood-based diagnostics, and
covariance structure analyses.

\subsubsection{Reconstruction of Individual Realizations}
\label{subsec:single-realization}

\begin{figure}[h!]
    \centering
    \includegraphics[width=1\linewidth]{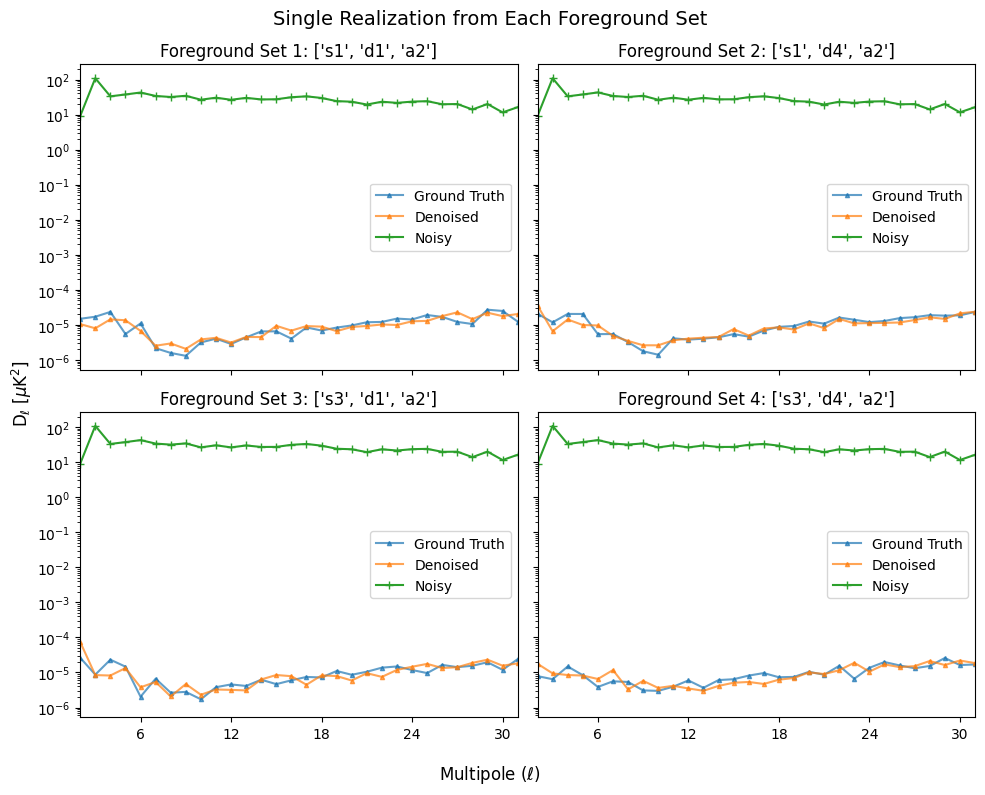}
    \caption{
Reconstruction of a single low-$\ell$ CMB $B$-mode realization for four distinct
foreground configurations (F1–F4; Section~\ref{subsec:testing-data}). In each panel,
the contaminated observed spectrum (green) is dominated by polarized foregrounds
and lensing across all multipoles, while the reverse VE--SDE reconstruction
(orange) successfully recovers the primordial $B$-mode spectrum (blue) without
explicit foreground modeling.
}
\label{fig:single_recon}
\end{figure}

We first assess the reconstruction efficiency at the level of individual
observational spectra. In Fig.\ref{fig:single_recon}, we show an example
realization from the test set containing primordial $B$-modes at $r=0.001$,
together with gravitational lensing, instrumental noise and polarized
foreground emission.

\noindent The observed spectrum (green) exhibits significant excess power across
multipoles $2 \le \ell \le 31$, such that the primordial signal is completely
subdominant.
\begin{figure}[h!]
    \centering
    \includegraphics[width=1\linewidth]{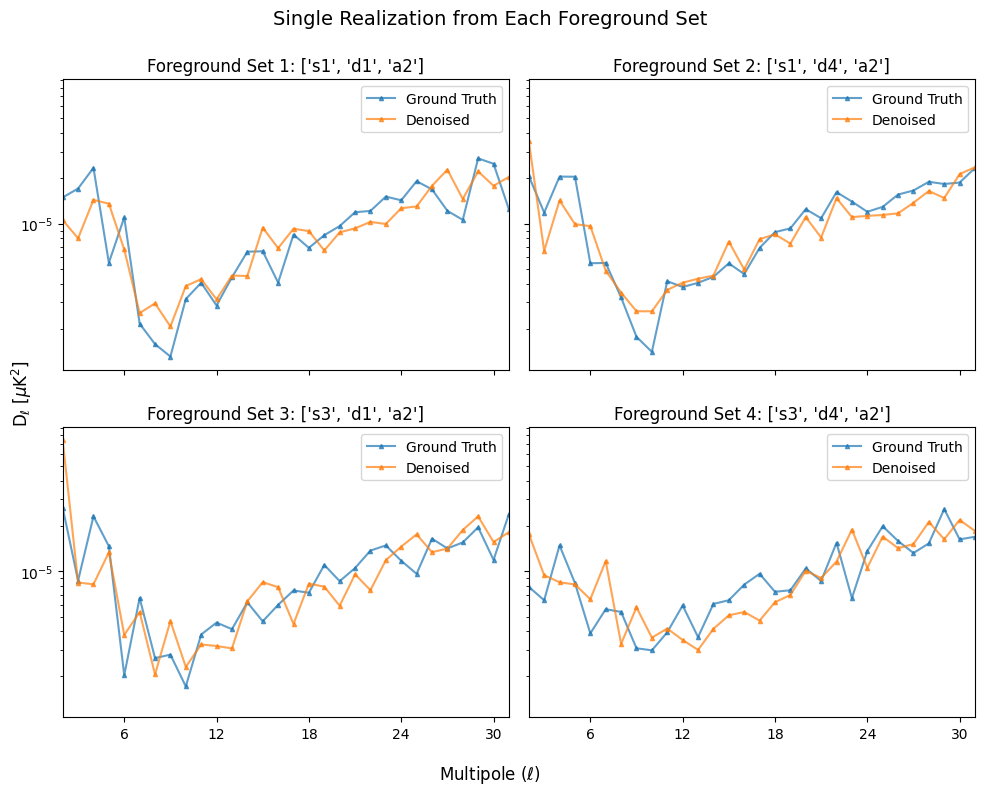}
   \caption{
Direct comparison of the reconstructed (orange) and ground-truth (blue)
primordial $B$-mode spectra for the same test realization shown in
Fig.~\ref{fig:single_recon}. The close agreement across all multipoles demonstrates that the reverse VE-SDE reconstruction accurately recovers both the amplitude and shape of the primordial signal.}
\label{fig:singlezoom}
\end{figure}Applying the reverse VE-SDE guided by the trained score model
successfully suppresses the contaminating components and recovers the
characteristic shape and amplitude of the true primordial spectrum. The reconstructed spectrum (orange) closely tracks the ground truth (blue)
across all multipoles, demonstrating that the learned score prior provides
sufficient structure to guide the reverse SDE to the correct solution, even
from a strongly contaminated observational input.
\begin{figure}[h!]
    \centering
    \includegraphics[width=1\linewidth]{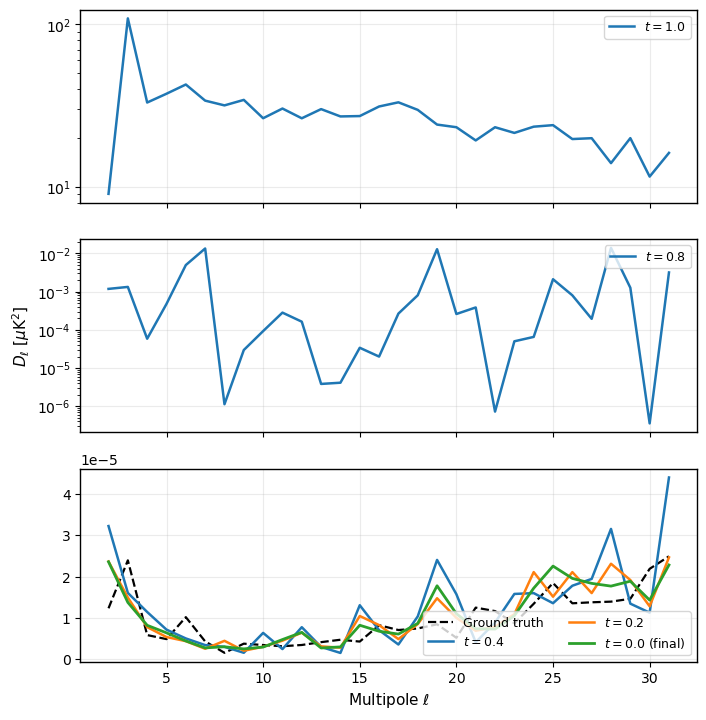}
    \caption{Evolution of the reconstructed $D_\ell$ spectrum for a single noisy observation
during reverse VE--SDE sampling.
\textit{Top panel:} Initial state at $t=1$, where the spectrum is dominated by noise and carries no primordial structure. \textit{Middle panel:} Intermediate diffusion stage ($t=0.8$), showing a gradual
reduction in amplitude and variance as stochastic noise is suppressed and the
spectrum begins to move toward the learned primordial distribution. \textit{Bottom panel:} Late-stage reconstruction at decreasing diffusion times
($t=0.4,\,0.2,\,0.0$), demonstrating convergence toward the primordial $B$-mode spectrum (black dashed line).}
\label{fig:reverse_sde_evolution}
\end{figure}
\begin{figure*}[b!]
\centering
\includegraphics[width=1\linewidth]{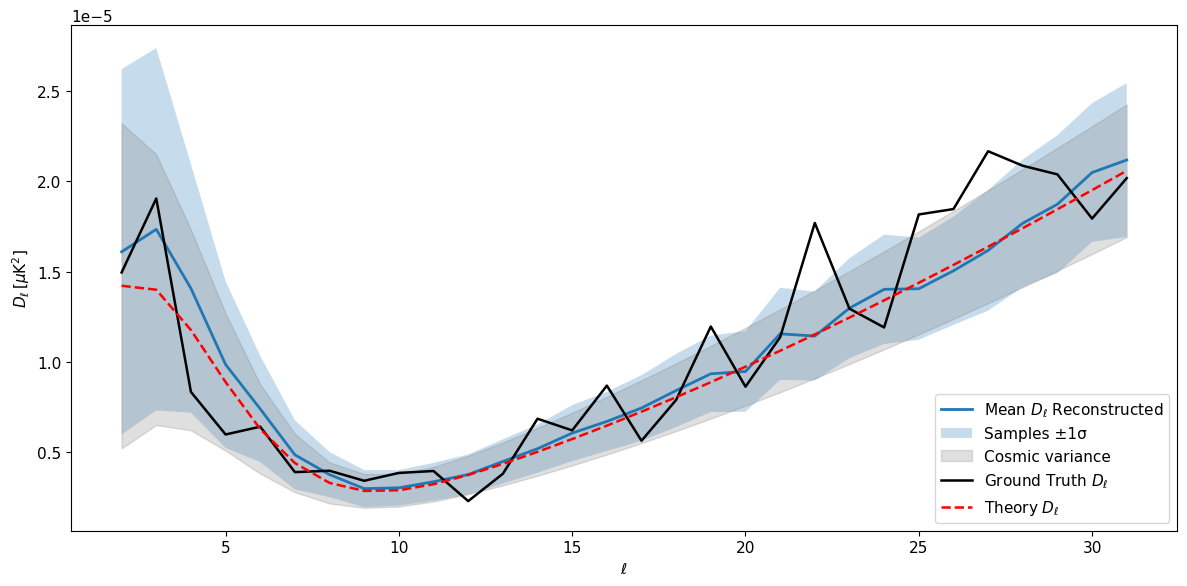}
\caption{
Ensemble-averaged reconstruction of the primordial $B$-mode $D_\ell$ spectrum
from $10{,}000$ independent reverse VE--SDE samples starting from the same
observed realization. The reconstructed mean (blue) closely follows both the
ground-truth spectrum (black) and the theoretical prediction (red dashed).
The shaded blue region denotes the $\pm1\sigma$ reconstruction uncertainty,
while the gray band indicates the expected cosmic-variance envelope. The
reconstruction uncertainty remains fully consistent with cosmic variance across
all multipoles, indicating unbiased and well-calibrated inference.
}
\label{fig:cv_compare}
\end{figure*}

\begin{figure}[h!]
    \centering
    \includegraphics[width=1\linewidth]{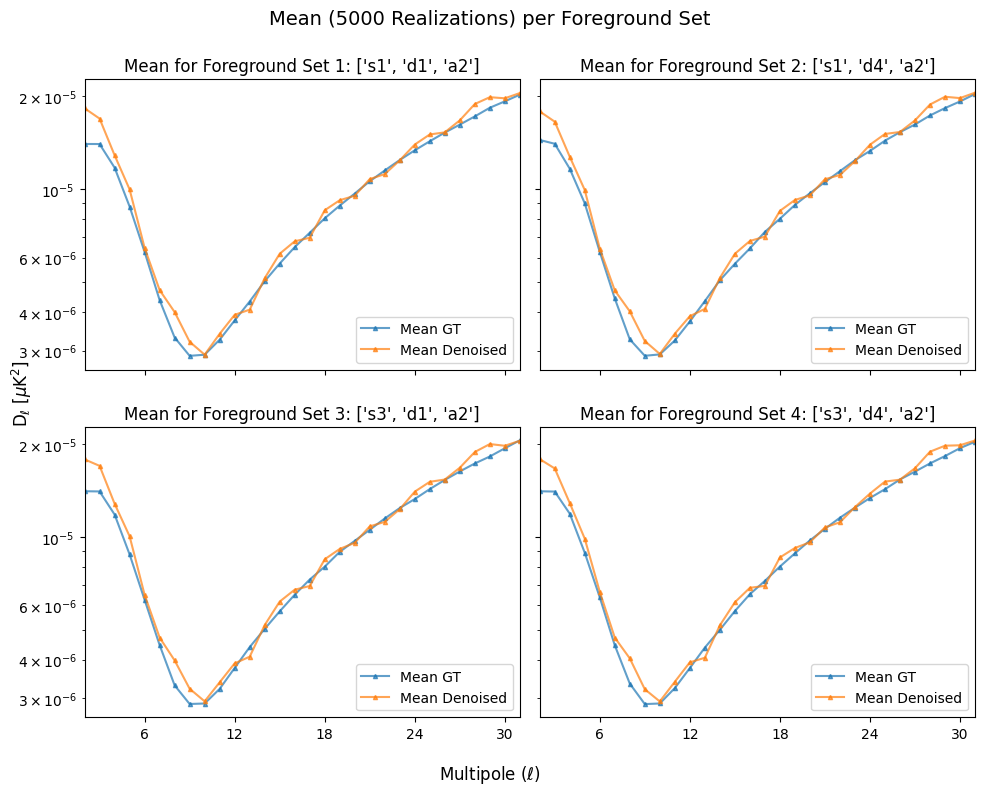} 
    \caption{
Ensemble-averaged primordial $B$-mode power spectrum obtained from $5{,}000$
independent realizations for each foreground configuration. The reconstructed mean (orange) closely matches the true primordial mean (blue) across all multipoles, demonstrating unbiased recovery in expectation and robustness of the reverse VE-SDE reconstruction to foreground complexity.}
\label{fig:ensemble_mean}
\end{figure}

\noindent To better visualize the accuracy of the denoised and delensed reconstruction,
Fig.\ref{fig:singlezoom} compares only the recovered spectrum to the ground
truth and the characteristic evolution of the spectrum during reverse SDE integration is shown in Fig.\ref{fig:reverse_sde_evolution}. While individual examples demonstrate strong match, it is
important to ensure that the method performs consistently across the full test
set. Fig.\ref{fig:ensemble_mean} shows the ensemble-averaged reconstructed
spectrum compared to the ensemble average of the true primordial spectra.
The near-perfect overlap indicates that the score-based model is unbiased in
the mean and does not systematically distort the recovered tensor power.

\subsubsection{Reconstruction Uncertainty and Cosmic Variance Consistency}
\label{subsec:ensemble}

To probe the stochastic properties of the reverse VE-SDE sampler and to assess
how reconstruction uncertainties compare with the fundamental cosmic-variance
limit, we perform $10{,}000$ independent reverse SDE reconstructions starting
from the \emph{same} observed spectrum. Each run differs only in the random
realization of the Wiener process driving the reverse-time dynamics, while the
trained score model and input observation are kept fixed. As shown in Fig.\ref{fig:cv_compare}, the mean of the reconstructed spectra
closely matches both the true primordial spectrum and the theoretical
prediction across all multipoles. This indicates that the stochastic nature of
the reverse SDE does not introduce any noticeable bias in the amplitude or
shape of the recovered signal. Moreover, the spread of the reconstructed
samples stays within the expected cosmic-variance range, meaning the method
does not produce uncertainties larger than the fundamental limits set by the
finite number of sky modes. This behavior has two practical advantages. First, multiple reverse-SDE
samples provide a reliable estimate of reconstruction uncertainty that is
consistent with the statistical limits of the problem. Second, the lack of
excess variance confirms that the learned score model is well-calibrated and
does not push solutions toward unrealistic regions of parameter space.

\subsubsection{Score field geometry validation}
\label{subsec:scorefield}

\begin{figure*}[t!]
    \centering
    \includegraphics[width=0.9\linewidth]{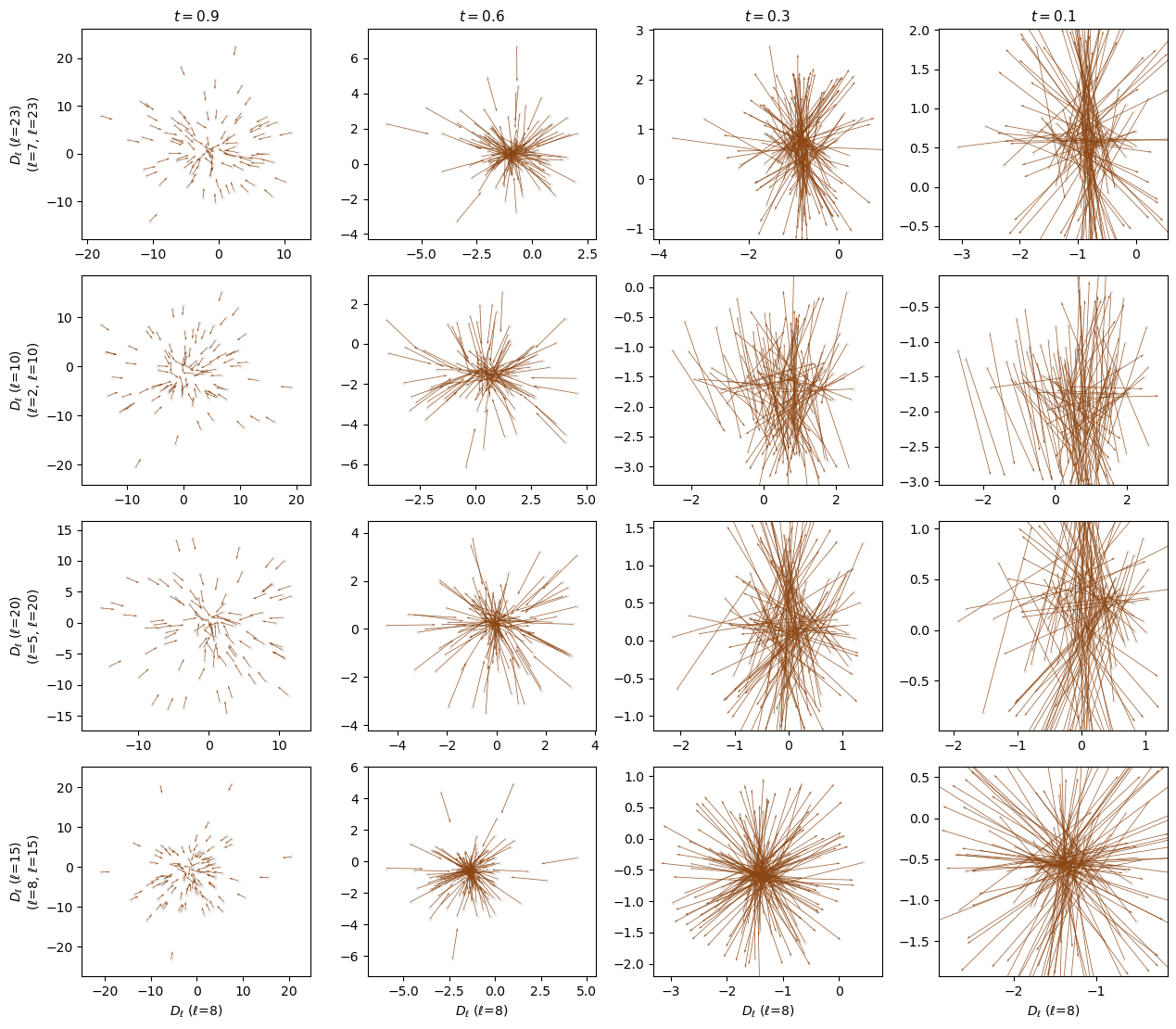}
\caption{
Visualization of the learned score field in $D_\ell$ space at multiple diffusion
times for selected multipole pairs.
\textit{Left to right:} Decreasing diffusion time $t=\{0.9,\,0.6,\,0.3,\,0.1\}$.
At large $t$, the score vectors exhibit an approximately isotropic pattern,
reflecting the near-Gaussian, noise-dominated distribution induced by the
forward diffusion process. As $t$ decreases, the vectors become increasingly
structured and aligned, indicating that the learned score consistently points
toward regions of higher data density. This evolution confirms that the network
has learned a stable and well-behaved approximation to the gradient of the
log-density required for accurate reverse VE--SDE sampling.
}

    \label{fig:score_vectors}
\end{figure*}
Since the quality of reverse SDE sampling depends on the accuracy of the learned score function, we inspect the geometry of the score field in log-normalized $D_\ell$ space at different diffusion times. In Fig.\ref{fig:score_vectors}, we visualize the score directions across the state space using quiver plots extracted from the trained model. At large diffusion times $t$ (high noise levels), the score vectors exhibit an approximately isotropic pattern, reflecting the fact that the perturbed distribution approaches a nearly Gaussian. As $t \rightarrow 0$, the vectors become increasingly structured and consistently point toward regions of higher data density, forming smooth trajectories that guide the reverse SDE toward the learned manifold of primordial spectra. This behavior is precisely what is expected from a well-trained estimator of the gradient of the log-probability density. The observed directional alignment at low noise indicates that the learned score provides stable and physically meaningful guidance for the sampling process, helping to ensure convergence to statistically consistent reconstructions of the primordial $B$-mode signal.

\begin{figure*}[t!]
    \centering
    \includegraphics[width=1\linewidth]{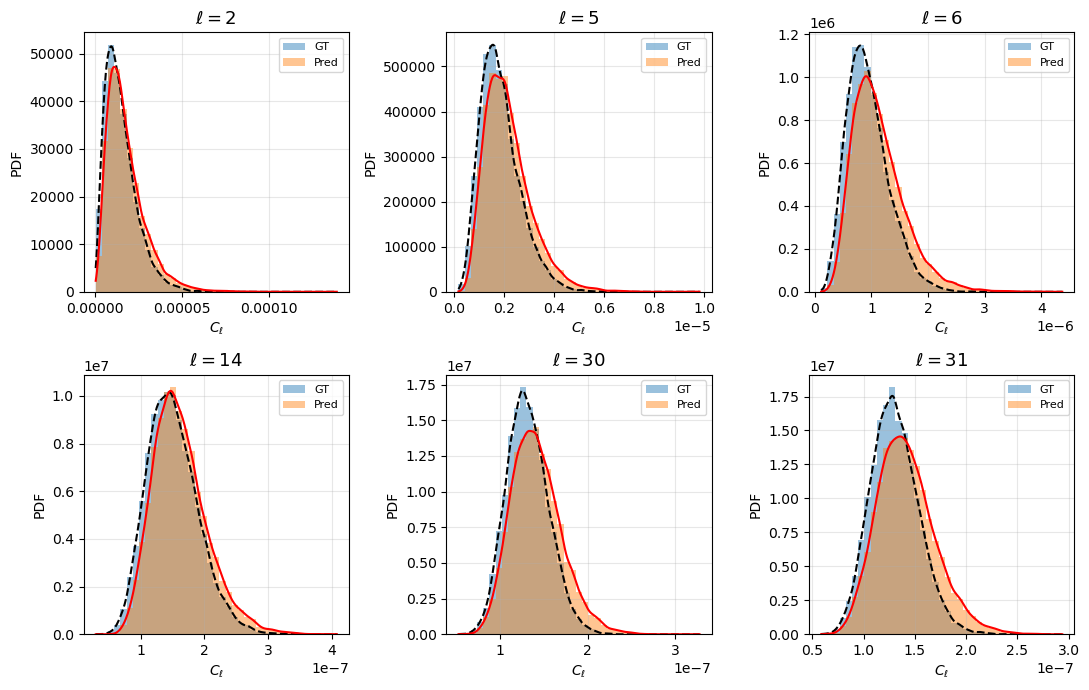}
    \caption{Marginal posterior distributions of the primordial $C_\ell$ for the multipoles
$\ell = \{2,\,5,\,6,\,14,\,30,\,31\}$, selected based on their comparatively larger KL divergence in Fig.~\ref{fig:kl_plot}. The reconstructed distributions (orange)
closely reproduce the ground-truth distributions (blue) in peak location and overall width, with small discrepancies appearing primarily as mild rightward shifts and enhanced high-$C_\ell$ tails. These tail differences dominate the KL
contribution and do not indicate a systematic bias in the bulk of the
distribution.}
    \label{fig:marginal_compare}
\end{figure*}




\subsubsection{Likelihood and Distribution Agreement}
\label{subsec:kl}
\begin{figure}[h!]
\centering
\includegraphics[width=1\linewidth]{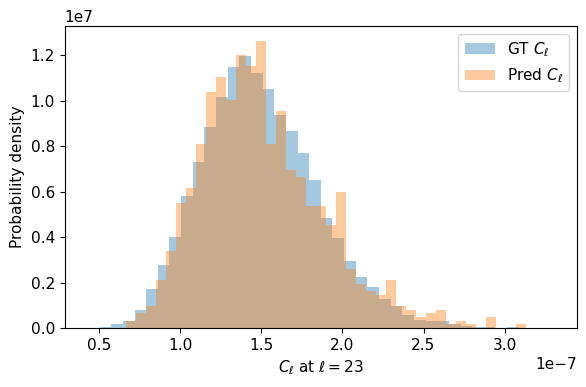}
\caption{
Histogram comparison of the marginal $C_\ell$ distributions for the ground truth
(blue) and reconstructed spectra (orange) at $\ell = 23$. The strong overlap in
both the central region and the distribution tails indicates that the
reconstruction accurately reproduces the underlying likelihood, including the
expected variance and non-Gaussian features at low multipoles.}
\label{fig:histo_l7}
\end{figure}

\begin{figure}[h!]
    \centering
    \includegraphics[width=1\linewidth]{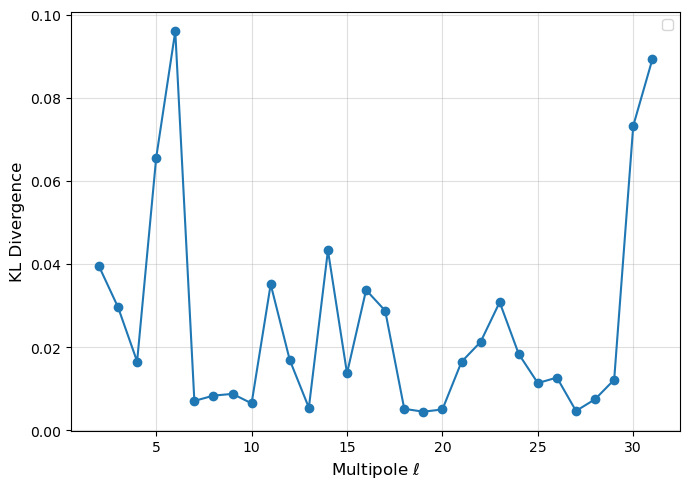}
    \caption{Kullback--Leibler (KL) divergence between the ground-truth and reconstructed
marginal $C_\ell$ distributions as a function of multipole $\ell$.
For the majority of multipoles, the KL divergence remains $\lesssim 0.04$,
indicating close agreement between the reconstructed and true likelihoods.}
\label{fig:kl_plot}
\end{figure}

To further evaluate how well the reconstructed spectra reproduce the underlying statistical distribution of the primordial signal, we compare both the marginal distribution shapes and the Kullback--Leibler (KL) divergence. Fig.\ref{fig:histo_l7} shows the marginal $C_\ell$ distribution comparison
at $\ell = 23$. The strong overlap in both the central region and the tails
indicates that the reconstructed spectra preserve realistic spread, skewness,
and non-Gaussian behavior expected at low multipoles.We further quantify this agreement using the KL divergence across multipoles,
as shown in Fig.\ref{fig:kl_plot}. For the majority of multipoles the KL divergence is $\lesssim 0.04$, and it remains below $0.1$ for all $\ell$. Only a few modes (for example around $\ell \approx 5$ and at the highest multipoles) show slightly elevated values, but these are still small enough that the reconstructed distributions closely track the ground–truth likelihood without introducing noticeable distortions.

For the multipoles where $\mathrm{KL}>0.04$, we compare the marginal reconstructed and ground-truth
$C_\ell$ distributions for the multipoles with the largest discrepancies,
namely $\ell = \{2,5,6,14,30,31\}$. As shown in
Fig.\ref{fig:marginal_compare}, the reconstructed posteriors remain well
aligned with the true distributions in both peak location and overall shape,
including skewness and tail behavior. The differences responsible for the KL
contribution arise primarily from small shifts in the high-variance tails
rather than from any systematic bias in the bulk of the distribution. This
confirms that the model faithfully reproduces the statistical properties of
the primordial $B$-mode signal even in the most challenging low-signal
regimes.

Together, these tests demonstrate that the method is accurate at the realization level, unbiased in the mean, uncertainty-calibrated relative to cosmic variance, and consistent with the ground-truth one-point likelihood across multipoles.

\section{Summary and Conclusion}
In this work, we developed a score-based stochastic differential equation (SDE)
framework that performs primordial $B$-mode reconstruction through a learned
reverse diffusion process in the low-multipole regime. By training a score model on simulated primordial spectra at $r=0.001$, the method learns the underlying data distribution and uses this learned score to iteratively remove noise, lensing contamination, and polarized foregrounds through the reverse VE-SDE. Because the model is trained only on primordial spectra and not on any specific foreground templates, the reconstruction is driven by the learned structure of the primordial signal rather than by assumptions about the nature of contaminants.

We demonstrated that the method accurately recovers both the amplitude and shape
of the primordial spectrum across a wide range of astrophysical conditions.
Tests performed on four realistic polarized-foreground models show that the
reconstructed spectra closely follow the ground truth without requiring any
explicit foreground modeling. An ensemble robustness study, involving 10,000
repeated reverse SDE runs on the same observation, confirms that the
reconstruction is unbiased, statistically stable, and does not introduce
additional variance beyond cosmic uncertainty. Furthermore,
distribution-level diagnostics including covariance matching, KL divergence,
and score-field directionality validate that high dimensional statistical
properties of the primordial signal are preserved. The generality of the score-based SDE framework makes it directly applicable to forthcoming CMB polarization missions. The same methodology can be retrained at different noise levels and frequency channels for experiments such as CMB-S4, the Simons Observatory, LiteBIRD, and ECHO. Moreover, because the model learns the score of the primordial distribution, it can also be used as a generative tool to simulate new realizations of the primordial $B$-mode spectrum consistent with a given tensor-to-scalar ratio. This dual capability reconstruction and high-fidelity simulation positions score-based SDEs as a powerful component of future CMB analysis pipelines.

Looking ahead, this work provides a foundation for developing more extensive models that learn the continuous family of primordial spectra over a range of $r$ values rather than a fixed amplitude. Such models could offer a robust, data-driven mechanism for probing the primordial gravitational-wave background and may eventually serve as state-of-the-art simulators for next-generation missions. With further refinement and broader training datasets, the score-based SDE framework has the potential to become a widely applicable tool for precision reconstruction of the primordial CMB $B$-mode signal.

\bibliography{references}

\begin{thebibliography}{71}
\providecommand{\natexlab}[1]{#1}
\providecommand{\url}[1]{\texttt{#1}}
\expandafter\ifx\csname urlstyle\endcsname\relax
  \providecommand{\doi}[1]{doi: #1}\else
  \providecommand{\doi}{doi: \begingroup \urlstyle{rm}\Url}\fi

\bibitem[Abazajian et~al.(2022)Abazajian, Addison, Adshead, Ahmed, Akerib, Ali, Allen, Alonso, Alvarez, Amin, et~al.]{abazajian2022cmb}
Kevork Abazajian, Graeme~E Addison, Peter Adshead, Zeeshan Ahmed, Daniel Akerib, Aamir Ali, Steven~W Allen, David Alonso, Marcelo Alvarez, Mustafa~A Amin, et~al.
\newblock Cmb-s4: Forecasting constraints on primordial gravitational waves.
\newblock \emph{The Astrophysical Journal}, 926\penalty0 (1):\penalty0 54, 2022.

\bibitem[Adak(2025)]{Adak:2025iyj}
Debabrata Adak.
\newblock {Deep Needlet: a CNN based full sky component separation method in Needlet space}.
\newblock \emph{JCAP}, 08:\penalty0 058, 2025.
\newblock \doi{10.1088/1475-7516/2025/08/058}.

\bibitem[Adak et~al.(2022)Adak, Sen, Basak, Delabrouille, Ghosh, Rotti, Martínez-Solaeche, and Souradeep]{Adak_2022}
Debabrata Adak, Aparajita Sen, Soumen Basak, Jacques Delabrouille, Tuhin Ghosh, Aditya Rotti, Ginés Martínez-Solaeche, and Tarun Souradeep.
\newblock B-mode forecast of cmb-bhārat.
\newblock \emph{Monthly Notices of the Royal Astronomical Society}, 514\penalty0 (2):\penalty0 3002–3016, May 2022.
\newblock ISSN 1365-2966.
\newblock \doi{10.1093/mnras/stac1474}.
\newblock URL \url{http://dx.doi.org/10.1093/mnras/stac1474}.

\bibitem[Aghanim et~al.(2020)Aghanim, Akrami, Ashdown, and et~al.]{2020}
N.~Aghanim, Y.~Akrami, M.~Ashdown, and et~al.
\newblock Planck2018 results: Vi. cosmological parameters.
\newblock \emph{Astronomy \&amp; Astrophysics}, 641:\penalty0 A6, September 2020.
\newblock ISSN 1432-0746.
\newblock \doi{10.1051/0004-6361/201833910}.
\newblock URL \url{http://dx.doi.org/10.1051/0004-6361/201833910}.

\bibitem[Anderson(1982)]{Anderson1982ReversetimeDE}
Brian. D.~O. Anderson.
\newblock Reverse-time diffusion equation models.
\newblock \emph{Stochastic Processes and their Applications}, 12:\penalty0 313--326, 1982.
\newblock URL \url{https://api.semanticscholar.org/CorpusID:3897405}.

\bibitem[Austermann et~al.(2012)Austermann, Aird, Beall, Becker, Bender, Benson, Bleem, Britton, Carlstrom, Chang, Chiang, Cho, Crawford, Crites, Datesman, de~Haan, Dobbs, George, Halverson, Harrington, Henning, Hilton, Holder, Holzapfel, Hoover, Huang, Hubmayr, Irwin, Keisler, Kennedy, Knox, Lee, Leitch, Li, Lueker, Marrone, McMahon, Mehl, Meyer, Montroy, Natoli, Nibarger, Niemack, Novosad, Padin, Pryke, Reichardt, Ruhl, Saliwanchik, Sayre, Schaffer, Shirokoff, Stark, Story, Vanderlinde, Vieira, Wang, Williamson, Yefremenko, Yoon, and Zahn]{Austermann_2012}
J.~E. Austermann, K.~A. Aird, J.~A. Beall, D.~Becker, A.~Bender, B.~A. Benson, L.~E. Bleem, J.~Britton, J.~E. Carlstrom, C.~L. Chang, H.~C. Chiang, H.-M. Cho, T.~M. Crawford, A.~T. Crites, A.~Datesman, T.~de~Haan, M.~A. Dobbs, E.~M. George, N.~W. Halverson, N.~Harrington, J.~W. Henning, G.~C. Hilton, G.~P. Holder, W.~L. Holzapfel, S.~Hoover, N.~Huang, J.~Hubmayr, K.~D. Irwin, R.~Keisler, J.~Kennedy, L.~Knox, A.~T. Lee, E.~Leitch, D.~Li, M.~Lueker, D.~P. Marrone, J.~J. McMahon, J.~Mehl, S.~S. Meyer, T.~E. Montroy, T.~Natoli, J.~P. Nibarger, M.~D. Niemack, V.~Novosad, S.~Padin, C.~Pryke, C.~L. Reichardt, J.~E. Ruhl, B.~R. Saliwanchik, J.~T. Sayre, K.~K. Schaffer, E.~Shirokoff, A.~A. Stark, K.~Story, K.~Vanderlinde, J.~D. Vieira, G.~Wang, R.~Williamson, V.~Yefremenko, K.~W. Yoon, and O.~Zahn.
\newblock Sptpol: an instrument for cmb polarization measurements with the south pole telescope.
\newblock In Wayne~S. Holland, editor, \emph{Millimeter, Submillimeter, and Far-Infrared Detectors and Instrumentation for Astronomy VI}. SPIE, September 2012.
\newblock \doi{10.1117/12.927286}.
\newblock URL \url{http://dx.doi.org/10.1117/12.927286}.

\bibitem[Bennett et~al.(2013)Bennett, Larson, Weiland, Jarosik, Hinshaw, Odegard, Smith, Hill, Gold, Halpern, Komatsu, Nolta, Page, Spergel, Wollack, Dunkley, Kogut, Limon, Meyer, Tucker, and Wright]{Bennett_2013}
C.~L. Bennett, D.~Larson, J.~L. Weiland, N.~Jarosik, G.~Hinshaw, N.~Odegard, K.~M. Smith, R.~S. Hill, B.~Gold, M.~Halpern, E.~Komatsu, M.~R. Nolta, L.~Page, D.~N. Spergel, E.~Wollack, J.~Dunkley, A.~Kogut, M.~Limon, S.~S. Meyer, G.~S. Tucker, and E.~L. Wright.
\newblock Nine-year wilkinson microwave anisotropy probe ( wmap ) observations: Final maps and results.
\newblock \emph{The Astrophysical Journal Supplement Series}, 208\penalty0 (2):\penalty0 20, September 2013.
\newblock ISSN 1538-4365.
\newblock \doi{10.1088/0067-0049/208/2/20}.
\newblock URL \url{http://dx.doi.org/10.1088/0067-0049/208/2/20}.

\bibitem[Bernardeau(1996)]{bernardeau1996weak}
F~Bernardeau.
\newblock Weak lensing detection in cmb maps.
\newblock \emph{arXiv preprint astro-ph/9611012}, 1996.

\bibitem[Blanchard and Schneider(1987)]{blanchard1987gravitational}
A~Blanchard and J~Schneider.
\newblock Gravitational lensing effect on the fluctuations of the cosmic background radiation.
\newblock \emph{Astronomy and Astrophysics (ISSN 0004-6361), vol. 184, no. 1-2, Oct. 1987, p. 1-6.}, 184:\penalty0 1--6, 1987.

\bibitem[Borrill et~al.(2025)Borrill, Clark, Delabrouille, Frolov, Ghosh, Hensley, Hicks, Krachmalnicoff, Lau, Norton, Pryke, Puglisi, Remazeilles, Russier, Thorne, Yao, and Zonca]{2025}
Julian Borrill, Susan~E. Clark, Jacques Delabrouille, Andrei~V. Frolov, Shamik Ghosh, Brandon~S. Hensley, Monica~D. Hicks, Nicoletta Krachmalnicoff, King Lau, Myra~M. Norton, Clement Pryke, Giuseppe Puglisi, Mathieu Remazeilles, Elisa Russier, Benjamin Thorne, Jian Yao, and Andrea Zonca.
\newblock Full-sky models of galactic microwave emission and polarization at subarcminute scales for the python sky model.
\newblock \emph{The Astrophysical Journal}, 991\penalty0 (1):\penalty0 23, September 2025.
\newblock ISSN 1538-4357.
\newblock \doi{10.3847/1538-4357/adf212}.
\newblock URL \url{http://dx.doi.org/10.3847/1538-4357/adf212}.

\bibitem[Boruah et~al.(2025)Boruah, Jacob, and Jain]{boruah2025diffusionbasedmassmapreconstruction}
Supranta~S. Boruah, Michael Jacob, and Bhuvnesh Jain.
\newblock Diffusion-based mass map reconstruction from weak lensing data, 2025.
\newblock URL \url{https://arxiv.org/abs/2502.04158}.

\bibitem[Caldeira et~al.(2019)Caldeira, Wu, Nord, Avestruz, Trivedi, and Story]{Caldeira:2018ojb}
Jo{\~a}o Caldeira, W.~L.~Kimmy Wu, Brian Nord, Camille Avestruz, Shubhendu Trivedi, and Kyle~T. Story.
\newblock {DeepCMB: Lensing Reconstruction of the Cosmic Microwave Background with Deep Neural Networks}.
\newblock \emph{Astron. Comput.}, 28:\penalty0 100307, 2019.
\newblock \doi{10.1016/j.ascom.2019.100307}.

\bibitem[Carones(2024)]{carones2024optimizingblindreconstructioncmb}
Alessandro Carones.
\newblock Optimizing blind reconstruction of cmb b-modes for future experiments, 2024.
\newblock URL \url{https://arxiv.org/abs/2406.15435}.

\bibitem[Chanda and Saha(2021)]{Chanda:2021qbf}
Pallav Chanda and Rajib Saha.
\newblock {An unbiased estimator of the full-sky CMB angular power spectrum at large scales using neural networks}.
\newblock \emph{Mon. Not. Roy. Astron. Soc.}, 508\penalty0 (3):\penalty0 4600--4609, 2021.
\newblock \doi{10.1093/mnras/stab2753}.

\bibitem[Collaboration et~al.(2023)Collaboration, Allys, Arnold, Aumont, Aurlien, Azzoni, Baccigalupi, Banday, Banerji, Barreiro, et~al.]{litebird2023probing}
LiteBIRD Collaboration, E~Allys, K~Arnold, J~Aumont, R~Aurlien, S~Azzoni, C~Baccigalupi, AJ~Banday, R~Banerji, RB~Barreiro, et~al.
\newblock Probing cosmic inflation with the litebird cosmic microwave background polarization survey.
\newblock \emph{Progress of Theoretical and Experimental Physics}, 2023\penalty0 (4):\penalty0 042F01, 2023.

\bibitem[{Delabrouille} et~al.(2013){Delabrouille}, {Betoule}, {Melin}, {Miville-Desch{\^e}nes}, {Gonzalez-Nuevo}, {Le Jeune}, {Castex}, {de Zotti}, {Basak}, {Ashdown}, {Aumont}, {Baccigalupi}, {Banday}, {Bernard}, {Bouchet}, {Clements}, {da Silva}, {Dickinson}, {Dodu}, {Dolag}, {Elsner}, {Fauvet}, {Fa{\"y}}, {Giardino}, {Leach}, {Lesgourgues}, {Liguori}, {Mac{\'\i}as-P{\'e}rez}, {Massardi}, {Matarrese}, {Mazzotta}, {Montier}, {Mottet}, {Paladini}, {Partridge}, {Piffaretti}, {Prezeau}, {Prunet}, {Ricciardi}, {Roman}, {Schaefer}, and {Toffolatti}]{2013AA553A96D}
J.~{Delabrouille}, M.~{Betoule}, J.-B. {Melin}, M.-A. {Miville-Desch{\^e}nes}, J.~{Gonzalez-Nuevo}, M.~{Le Jeune}, G.~{Castex}, G.~{de Zotti}, S.~{Basak}, M.~{Ashdown}, J.~{Aumont}, C.~{Baccigalupi}, A.~J. {Banday}, J.-P. {Bernard}, F.~R. {Bouchet}, D.~L. {Clements}, A.~{da Silva}, C.~{Dickinson}, F.~{Dodu}, K.~{Dolag}, F.~{Elsner}, L.~{Fauvet}, G.~{Fa{\"y}}, G.~{Giardino}, S.~{Leach}, J.~{Lesgourgues}, M.~{Liguori}, J.~F. {Mac{\'\i}as-P{\'e}rez}, M.~{Massardi}, S.~{Matarrese}, P.~{Mazzotta}, L.~{Montier}, S.~{Mottet}, R.~{Paladini}, B.~{Partridge}, R.~{Piffaretti}, G.~{Prezeau}, S.~{Prunet}, S.~{Ricciardi}, M.~{Roman}, B.~{Schaefer}, and L.~{Toffolatti}.
\newblock {The pre-launch Planck Sky Model: a model of sky emission at submillimetre to centimetre wavelengths}.
\newblock \emph{\aap}, 553:\penalty0 A96, May 2013.
\newblock \doi{10.1051/0004-6361/201220019}.

\bibitem[Dobler and Finkbeiner(2008)]{Dobler_2008}
Gregory Dobler and Douglas~P. Finkbeiner.
\newblock Extended anomalous foreground emission in thewmapthree‐year data.
\newblock \emph{The Astrophysical Journal}, 680\penalty0 (2):\penalty0 1222–1234, June 2008.
\newblock ISSN 1538-4357.
\newblock \doi{10.1086/587862}.
\newblock URL \url{http://dx.doi.org/10.1086/587862}.

\bibitem[Eriksen et~al.(2008)Eriksen, Jewell, Dickinson, Banday, Górski, and Lawrence]{Eriksen_2008}
H.~K. Eriksen, J.~B. Jewell, C.~Dickinson, A.~J. Banday, K.~M. Górski, and C.~R. Lawrence.
\newblock Joint bayesian component separation and cmb power spectrum estimation.
\newblock \emph{The Astrophysical Journal}, 676\penalty0 (1):\penalty0 10–32, March 2008.
\newblock ISSN 1538-4357.
\newblock \doi{10.1086/525277}.
\newblock URL \url{http://dx.doi.org/10.1086/525277}.

\bibitem[{Farsian} et~al.(2020){Farsian}, {Krachmalnicoff}, and {Baccigalupi}]{2020JCAP...07..017F}
F.~{Farsian}, N.~{Krachmalnicoff}, and C.~{Baccigalupi}.
\newblock {Foreground model recognition through Neural Networks for CMB B-mode observations}.
\newblock \emph{\jcap}, 2020\penalty0 (7):\penalty0 017, July 2020.
\newblock \doi{10.1088/1475-7516/2020/07/017}.

\bibitem[{Finkbeiner} et~al.(1999){Finkbeiner}, {Davis}, and {Schlegel}]{1999ApJ...524..867F}
Douglas~P. {Finkbeiner}, Marc {Davis}, and David~J. {Schlegel}.
\newblock {Extrapolation of Galactic Dust Emission at 100 Microns to Cosmic Microwave Background Radiation Frequencies Using FIRAS}.
\newblock \emph{\apj}, 524\penalty0 (2):\penalty0 867--886, October 1999.
\newblock \doi{10.1086/307852}.

\bibitem[Flöss et~al.(2024)Flöss, Coulton, Duivenvoorden, Villaescusa-Navarro, and Wandelt]{flöss2024denoisingdiffusiondelensingdelight}
Thomas Flöss, William~R. Coulton, Adriaan~J. Duivenvoorden, Francisco Villaescusa-Navarro, and Benjamin~D. Wandelt.
\newblock Denoising diffusion delensing delight: Reconstructing the non-gaussian cmb lensing potential with diffusion models, 2024.
\newblock URL \url{https://arxiv.org/abs/2405.05598}.

\bibitem[Gorski et~al.(2005)Gorski, Hivon, Banday, Wandelt, Hansen, Reinecke, and Bartelmann]{Gorski_2005}
K.~M. Gorski, E.~Hivon, A.~J. Banday, B.~D. Wandelt, F.~K. Hansen, M.~Reinecke, and M.~Bartelmann.
\newblock Healpix: A framework for high‐resolution discretization and fast analysis of data distributed on the sphere.
\newblock \emph{The Astrophysical Journal}, 622\penalty0 (2):\penalty0 759–771, April 2005.
\newblock ISSN 1538-4357.
\newblock \doi{10.1086/427976}.
\newblock URL \url{http://dx.doi.org/10.1086/427976}.

\bibitem[Group et~al.(2025)Group, Borrill, Clark, Delabrouille, Frolov, Ghosh, Hensley, Hicks, Krachmalnicoff, Lau, Norton, Pryke, Puglisi, Remazeilles, Russier, Thorne, Yao, and Zonca]{Panexp_2025}
The Pan-Experiment Galactic~Science Group, Julian Borrill, Susan~E. Clark, Jacques Delabrouille, Andrei~V. Frolov, Shamik Ghosh, Brandon~S. Hensley, Monica~D. Hicks, Nicoletta Krachmalnicoff, King Lau, Myra~M. Norton, Clement Pryke, Giuseppe Puglisi, Mathieu Remazeilles, Elisa Russier, Benjamin Thorne, Jian Yao, and Andrea Zonca.
\newblock Full-sky models of galactic microwave emission and polarization at subarcminute scales for the python sky model.
\newblock \emph{The Astrophysical Journal}, 991\penalty0 (1):\penalty0 23, 2025.
\newblock \doi{10.3847/1538-4357/adf212}.
\newblock URL \url{https://iopscience.iop.org/article/10.3847/1538-4357/adf212}.

\bibitem[Guth(1981)]{PhysRevD.23.347}
Alan~H. Guth.
\newblock Inflationary universe: A possible solution to the horizon and flatness problems.
\newblock \emph{Phys. Rev. D}, 23:\penalty0 347--356, Jan 1981.
\newblock \doi{10.1103/PhysRevD.23.347}.
\newblock URL \url{https://link.aps.org/doi/10.1103/PhysRevD.23.347}.

\bibitem[Guth and Pi(1985)]{PhysRevD.32.1899}
Alan~H. Guth and So-Young Pi.
\newblock Quantum mechanics of the scalar field in the new inflationary universe.
\newblock \emph{Phys. Rev. D}, 32:\penalty0 1899--1920, Oct 1985.
\newblock \doi{10.1103/PhysRevD.32.1899}.
\newblock URL \url{https://link.aps.org/doi/10.1103/PhysRevD.32.1899}.

\bibitem[Hawking(1982)]{hawking1982development}
Stephen~W Hawking.
\newblock The development of irregularities in a single bubble inflationary universe.
\newblock \emph{Physics Letters B}, 115\penalty0 (4):\penalty0 295--297, 1982.

\bibitem[Joseph et~al.(2023)Joseph, Purkayastha, and Saha]{Joseph_2023}
Albin Joseph, Ujjal Purkayastha, and Rajib Saha.
\newblock A foreground model-independent bayesian cmb temperature and polarization signal reconstruction and cosmological parameter estimation over large angular scales.
\newblock \emph{Monthly Notices of the Royal Astronomical Society}, 520\penalty0 (1):\penalty0 976–987, January 2023.
\newblock ISSN 1365-2966.
\newblock \doi{10.1093/mnras/stad187}.
\newblock URL \url{http://dx.doi.org/10.1093/mnras/stad187}.

\bibitem[Kamionkowski and Kovetz(2016)]{Kamionkowski_2016}
Marc Kamionkowski and Ely~D. Kovetz.
\newblock The quest for b modes from inflationary gravitational waves.
\newblock \emph{Annual Review of Astronomy and Astrophysics}, 54\penalty0 (1):\penalty0 227–269, September 2016.
\newblock ISSN 1545-4282.
\newblock \doi{10.1146/annurev-astro-081915-023433}.
\newblock URL \url{http://dx.doi.org/10.1146/annurev-astro-081915-023433}.

\bibitem[Kermish et~al.(2012)Kermish, Ade, Anthony, Arnold, Barron, Boettger, Borrill, Chapman, Chinone, Dobbs, Errard, Fabbian, Flanigan, Fuller, Ghribi, Grainger, Halverson, Hasegawa, Hattori, Hazumi, Holzapfel, Howard, Hyland, Jaffe, Keating, Kisner, Lee, Le~Jeune, Linder, Lungu, Matsuda, Matsumura, Meng, Miller, Morii, Moyerman, Myers, Nishino, Paar, Quealy, Reichardt, Richards, Ross, Shimizu, Shimon, Shimmin, Sholl, Siritanasak, Spieler, Stebor, Steinbach, Stompor, Suzuki, Tomaru, Tucker, and Zahn]{Kermish_2012}
Zigmund~D. Kermish, Peter Ade, Aubra Anthony, Kam Arnold, Darcy Barron, David Boettger, Julian Borrill, Scott Chapman, Yuji Chinone, Matt~A. Dobbs, Josquin Errard, Giulio Fabbian, Daniel Flanigan, George Fuller, Adnan Ghribi, Will Grainger, Nils Halverson, Masaya Hasegawa, Kaori Hattori, Masashi Hazumi, William~L. Holzapfel, Jacob Howard, Peter Hyland, Andrew Jaffe, Brian Keating, Theodore Kisner, Adrian~T. Lee, Maude Le~Jeune, Eric Linder, Marius Lungu, Frederick Matsuda, Tomotake Matsumura, Xiaofan Meng, Nathan~J. Miller, Hideki Morii, Stephanie Moyerman, Mike~J. Myers, Haruki Nishino, Hans Paar, Erin Quealy, Christian~L. Reichardt, Paul~L. Richards, Colin Ross, Akie Shimizu, Meir Shimon, Chase Shimmin, Mike Sholl, Praween Siritanasak, Helmuth Spieler, Nathan Stebor, Bryan Steinbach, Radek Stompor, Aritoki Suzuki, Takayuki Tomaru, Carole Tucker, and Oliver Zahn.
\newblock The polarbear experiment.
\newblock In Wayne~S. Holland, editor, \emph{Millimeter, Submillimeter, and Far-Infrared Detectors and Instrumentation for Astronomy VI}. SPIE, September 2012.
\newblock \doi{10.1117/12.926354}.
\newblock URL \url{http://dx.doi.org/10.1117/12.926354}.

\bibitem[Kingma and Ba(2017)]{kingma2017adammethodstochasticoptimization}
Diederik~P. Kingma and Jimmy Ba.
\newblock Adam: A method for stochastic optimization, 2017.
\newblock URL \url{https://arxiv.org/abs/1412.6980}.

\bibitem[{Kogut}(2012)]{2012ApJ...753..110K}
A.~{Kogut}.
\newblock {Synchrotron Spectral Curvature from 22 MHz to 23 GHz}.
\newblock \emph{\apj}, 753\penalty0 (2):\penalty0 110, July 2012.
\newblock \doi{10.1088/0004-637X/753/2/110}.

\bibitem[Kogut(2012)]{Kogut_2012}
A.~Kogut.
\newblock Synchrotron spectral curvature from 22 mhz to 23 ghz.
\newblock \emph{The Astrophysical Journal}, 753\penalty0 (2):\penalty0 110, June 2012.
\newblock ISSN 1538-4357.
\newblock \doi{10.1088/0004-637x/753/2/110}.
\newblock URL \url{http://dx.doi.org/10.1088/0004-637X/753/2/110}.

\bibitem[Lewis et~al.(2000)Lewis, Challinor, and Lasenby]{Lewis_2000}
Antony Lewis, Anthony Challinor, and Anthony Lasenby.
\newblock Efficient computation of cosmic microwave background anisotropies in closed friedmann‐robertson‐walker models.
\newblock \emph{The Astrophysical Journal}, 538\penalty0 (2):\penalty0 473–476, August 2000.
\newblock ISSN 1538-4357.
\newblock \doi{10.1086/309179}.
\newblock URL \url{http://dx.doi.org/10.1086/309179}.

\bibitem[Linde(1982)]{Linde:1981mu}
Andrei~D. Linde.
\newblock {A New Inflationary Universe Scenario: A Possible Solution of the Horizon, Flatness, Homogeneity, Isotropy and Primordial Monopole Problems}.
\newblock \emph{Phys. Lett. B}, 108:\penalty0 389--393, 1982.
\newblock \doi{10.1016/0370-2693(82)91219-9}.

\bibitem[Lizarraga et~al.(2025)Lizarraga, Jiang, Nowack, Li, Wu, Boscoe, and Do]{lizarraga2025understandinggalaxymorphologyevolution}
Andrew Lizarraga, Eric~Hanchen Jiang, Jacob Nowack, Yun~Qi Li, Ying~Nian Wu, Bernie Boscoe, and Tuan Do.
\newblock Understanding galaxy morphology evolution through cosmic time via redshift conditioned diffusion models, 2025.
\newblock URL \url{https://arxiv.org/abs/2411.18440}.

\bibitem[M et~al.(2025)M, Jaiswal, Das, and Parattu]{m2025deeplearningcmbforeground}
Obasho M, Shambhavi Jaiswal, Santanu Das, and Krishna~Mohan Parattu.
\newblock Deep learning for cmb foreground removal and beam deconvolution: A u-net gan approach, 2025.
\newblock URL \url{https://arxiv.org/abs/2509.00139}.

\bibitem[{Miville-Desch{\^e}nes} et~al.(2008){Miville-Desch{\^e}nes}, {Ysard}, {Lavabre}, {Ponthieu}, {Mac{\'\i}as-P{\'e}rez}, {Aumont}, and {Bernard}]{2008AA4901093M}
M.-A. {Miville-Desch{\^e}nes}, N.~{Ysard}, A.~{Lavabre}, N.~{Ponthieu}, J.~F. {Mac{\'\i}as-P{\'e}rez}, J.~{Aumont}, and J.~P. {Bernard}.
\newblock {Separation of anomalous and synchrotron emissions using WMAP polarization data}.
\newblock \emph{\aap}, 490\penalty0 (3):\penalty0 1093--1102, November 2008.
\newblock \doi{10.1051/0004-6361:200809484}.

\bibitem[Naess et~al.(2014)Naess, Hasselfield, McMahon, Niemack, Addison, Ade, Allison, Amiri, Battaglia, Beall, de~Bernardis, Bond, Britton, Calabrese, Cho, Coughlin, Crichton, Das, Datta, Devlin, Dicker, Dunkley, Dünner, Fowler, Fox, Gallardo, Grace, Gralla, Hajian, Halpern, Henderson, Hill, Hilton, Hilton, Hincks, Hlozek, Ho, Hubmayr, Huffenberger, Hughes, Infante, Irwin, Jackson, Kasanda, Klein, Koopman, Kosowsky, Li, Louis, Lungu, Madhavacheril, Marriage, Maurin, Menanteau, Moodley, Munson, Newburgh, Nibarger, Nolta, Page, Pappas, Partridge, Rojas, Schmitt, Sehgal, Sherwin, Sievers, Simon, Spergel, Staggs, Switzer, Thornton, Trac, Tucker, Uehara, Engelen, Ward, and Wollack]{Naess_2014}
Sigurd Naess, Matthew Hasselfield, Jeff McMahon, Michael~D. Niemack, Graeme~E. Addison, Peter A.~R. Ade, Rupert Allison, Mandana Amiri, Nick Battaglia, James~A. Beall, Francesco de~Bernardis, J~Richard Bond, Joe Britton, Erminia Calabrese, Hsiao-mei Cho, Kevin Coughlin, Devin Crichton, Sudeep Das, Rahul Datta, Mark~J. Devlin, Simon~R. Dicker, Joanna Dunkley, Rolando Dünner, Joseph~W. Fowler, Anna~E. Fox, Patricio Gallardo, Emily Grace, Megan Gralla, Amir Hajian, Mark Halpern, Shawn Henderson, J.~Colin Hill, Gene~C. Hilton, Matt Hilton, Adam~D. Hincks, Renée Hlozek, Patty Ho, Johannes Hubmayr, Kevin~M. Huffenberger, John~P. Hughes, Leopoldo Infante, Kent Irwin, Rebecca Jackson, Simon~Muya Kasanda, Jeff Klein, Brian Koopman, Arthur Kosowsky, Dale Li, Thibaut Louis, Marius Lungu, Mathew Madhavacheril, Tobias~A. Marriage, Loïc Maurin, Felipe Menanteau, Kavilan Moodley, Charles Munson, Laura Newburgh, John Nibarger, Michael~R. Nolta, Lyman~A. Page, Christine Pappas, Bruce Partridge, Felipe Rojas, Benjamin~L.
  Schmitt, Neelima Sehgal, Blake~D. Sherwin, Jon Sievers, Sara Simon, David~N. Spergel, Suzanne~T. Staggs, Eric~R. Switzer, Robert Thornton, Hy~Trac, Carole Tucker, Masao Uehara, Alexander~Van Engelen, Jonathan~T. Ward, and Edward~J. Wollack.
\newblock The atacama cosmology telescope: Cmb polarization at 200 \&lt; l \&lt; 9000.
\newblock \emph{Journal of Cosmology and Astroparticle Physics}, 2014\penalty0 (10):\penalty0 007–007, October 2014.
\newblock ISSN 1475-7516.
\newblock \doi{10.1088/1475-7516/2014/10/007}.
\newblock URL \url{http://dx.doi.org/10.1088/1475-7516/2014/10/007}.

\bibitem[Namikawa et~al.(2022)Namikawa, Baleato~Lizancos, Robertson, Sherwin, Challinor, Alonso, Azzoni, Baccigalupi, Calabrese, Carron, et~al.]{namikawa2022simons}
Toshiya Namikawa, Anton Baleato~Lizancos, Naomi Robertson, Blake~D Sherwin, Anthony Challinor, David Alonso, Susanna Azzoni, Carlo Baccigalupi, Erminia Calabrese, Julien Carron, et~al.
\newblock Simons observatory: Constraining inflationary gravitational waves with multitracer b-mode delensing.
\newblock \emph{Physical Review D}, 105\penalty0 (2):\penalty0 023511, 2022.

\bibitem[O'Dea et~al.(2007)O'Dea, Challinor, and Johnson]{o2007systematic}
Daniel O'Dea, Anthony Challinor, and Bradley~R Johnson.
\newblock Systematic errors in cosmic microwave background polarization measurements.
\newblock \emph{Monthly Notices of the Royal Astronomical Society}, 376\penalty0 (4):\penalty0 1767--1783, 2007.

\bibitem[Pal et~al.(2024)Pal, Yadav, Saha, and Souradeep]{pal2024accurateunbiasedreconstructioncmb}
Srikanta Pal, Sarvesh~Kumar Yadav, Rajib Saha, and Tarun Souradeep.
\newblock Accurate and unbiased reconstruction of cmb b mode using deep learning, 2024.
\newblock URL \url{https://arxiv.org/abs/2404.18100}.

\bibitem[Paoletti et~al.(2022)Paoletti, Finelli, Valiviita, and Hazumi]{paoletti2022planck}
Daniela Paoletti, Fabio Finelli, Jussi Valiviita, and Masashi Hazumi.
\newblock Planck and bicep/keck array 2018 constraints on primordial gravitational waves and perspectives for future b-mode polarization measurements.
\newblock \emph{Physical Review D}, 106\penalty0 (8):\penalty0 083528, 2022.

\bibitem[{Peebles}(1993)]{1993ppc..book.....P}
P.~J.~E. {Peebles}.
\newblock \emph{{Principles of Physical Cosmology}}.
\newblock 1993.
\newblock \doi{10.1515/9780691206721}.

\bibitem[Peebles and Ratra(2003)]{Peebles_2003}
P.~J.~E. Peebles and Bharat Ratra.
\newblock The cosmological constant and dark energy.
\newblock \emph{Reviews of Modern Physics}, 75\penalty0 (2):\penalty0 559–606, April 2003.
\newblock ISSN 1539-0756.
\newblock \doi{10.1103/revmodphys.75.559}.
\newblock URL \url{http://dx.doi.org/10.1103/RevModPhys.75.559}.

\bibitem[Prunet and Lazarian(1999)]{Prunet:1999mh}
S.~Prunet and A.~Lazarian.
\newblock {Polarized foreground from thermal dust emission}.
\newblock 2 1999.

\bibitem[Ramachandran et~al.(2017)Ramachandran, Zoph, and Le]{ramachandran2017searchingactivationfunctions}
Prajit Ramachandran, Barret Zoph, and Quoc~V. Le.
\newblock Searching for activation functions, 2017.
\newblock URL \url{https://arxiv.org/abs/1710.05941}.

\bibitem[Ratra and Peebles(1988)]{ratra1988cosmological}
Bharat Ratra and Philip~JE Peebles.
\newblock Cosmological consequences of a rolling homogeneous scalar field.
\newblock \emph{Physical Review D}, 37\penalty0 (12):\penalty0 3406, 1988.

\bibitem[{Remazeilles} et~al.(2015){Remazeilles}, {Dickinson}, {Banday}, {Bigot-Sazy}, and {Ghosh}]{2015MNRAS.451.4311R}
M.~{Remazeilles}, C.~{Dickinson}, A.~J. {Banday}, M.-A. {Bigot-Sazy}, and T.~{Ghosh}.
\newblock {An improved source-subtracted and destriped 408-MHz all-sky map}.
\newblock \emph{\mnras}, 451\penalty0 (4):\penalty0 4311--4327, August 2015.
\newblock \doi{10.1093/mnras/stv1274}.

\bibitem[Remazeilles et~al.(2018)Remazeilles, Banday, Baccigalupi, Basak, Bonaldi, Zotti, Delabrouille, Dickinson, Eriksen, Errard, Fernandez-Cobos, Fuskeland, Hervías-Caimapo, López-Caniego, Martinez-González, Roman, Vielva, Wehus, Achucarro, Ade, Allison, Ashdown, Ballardini, Banerji, Bartlett, Bartolo, Baumann, Bersanelli, Bonato, Borrill, Bouchet, Boulanger, Brinckmann, Bucher, Burigana, Buzzelli, Cai, Calvo, Carvalho, Castellano, Challinor, Chluba, Clesse, Colantoni, Coppolecchia, Crook, D’Alessandro, de~Bernardis, de~Gasperis, Diego, Valentino, Feeney, Ferraro, Finelli, Forastieri, Galli, Genova-Santos, Gerbino, González-Nuevo, Grandis, Greenslade, Hagstotz, Hanany, Handley, Hernandez-Monteagudo, Hills, Hivon, Kiiveri, Kisner, Kitching, Kunz, Kurki-Suonio, Lamagna, Lasenby, Lattanzi, Lesgourgues, Lewis, Liguori, Lindholm, Luzzi, Maffei, Martins, Masi, Matarrese, McCarthy, Melin, Melchiorri, Molinari, Monfardini, Natoli, Negrello, Notari, Paiella, Paoletti, Patanchon, Piat, Pisano, Polastri,
  Polenta, Pollo, Poulin, Quartin, Rubino-Martin, Salvati, Tartari, Tomasi, Tramonte, Trappe, Trombetti, Tucker, Valiviita, de~Weijgaert, Tent, Vennin, Vittorio, Young, and Zannoni]{Remazeilles_2018}
M.~Remazeilles, A.J. Banday, C.~Baccigalupi, S.~Basak, A.~Bonaldi, G.~De Zotti, J.~Delabrouille, C.~Dickinson, H.~K. Eriksen, J.~Errard, R.~Fernandez-Cobos, U.~Fuskeland, C.~Hervías-Caimapo, M.~López-Caniego, E.~Martinez-González, M.~Roman, P.~Vielva, I.~Wehus, A.~Achucarro, P.~Ade, R.~Allison, M.~Ashdown, M.~Ballardini, R.~Banerji, J.~Bartlett, N.~Bartolo, D.~Baumann, M.~Bersanelli, M.~Bonato, J.~Borrill, F.~Bouchet, F.~Boulanger, T.~Brinckmann, M.~Bucher, C.~Burigana, A.~Buzzelli, Z.-Y. Cai, M.~Calvo, C.-S. Carvalho, G.~Castellano, A.~Challinor, J.~Chluba, S.~Clesse, I.~Colantoni, A.~Coppolecchia, M.~Crook, G.~D’Alessandro, P.~de~Bernardis, G.~de~Gasperis, J.-M. Diego, E.~Di Valentino, S.~Feeney, S.~Ferraro, F.~Finelli, F.~Forastieri, S.~Galli, R.~Genova-Santos, M.~Gerbino, J.~González-Nuevo, S.~Grandis, J.~Greenslade, S.~Hagstotz, S.~Hanany, W.~Handley, C.~Hernandez-Monteagudo, M.~Hills, E.~Hivon, K.~Kiiveri, T.~Kisner, T.~Kitching, M.~Kunz, H.~Kurki-Suonio, L.~Lamagna, A.~Lasenby, M.~Lattanzi,
  J.~Lesgourgues, A.~Lewis, M.~Liguori, V.~Lindholm, G.~Luzzi, B.~Maffei, C.J.A.P. Martins, S.~Masi, S.~Matarrese, D.~McCarthy, J.-B. Melin, A.~Melchiorri, D.~Molinari, A.~Monfardini, P.~Natoli, M.~Negrello, A.~Notari, A.~Paiella, D.~Paoletti, G.~Patanchon, M.~Piat, G.~Pisano, L.~Polastri, G.~Polenta, A.~Pollo, V.~Poulin, M.~Quartin, J.-A. Rubino-Martin, L.~Salvati, A.~Tartari, M.~Tomasi, D.~Tramonte, N.~Trappe, T.~Trombetti, C.~Tucker, J.~Valiviita, R.~Van de~Weijgaert, B.~van Tent, V.~Vennin, N.~Vittorio, K.~Young, and M.~Zannoni.
\newblock Exploring cosmic origins with core:b-mode component separation.
\newblock \emph{Journal of Cosmology and Astroparticle Physics}, 2018\penalty0 (04):\penalty0 023–023, April 2018.
\newblock ISSN 1475-7516.
\newblock \doi{10.1088/1475-7516/2018/04/023}.
\newblock URL \url{http://dx.doi.org/10.1088/1475-7516/2018/04/023}.

\bibitem[Riveros et~al.(2025)Riveros, Saavedra, Hortua, Garcia~Farieta, and Olier]{Riveros_2025}
Julieth~K. Riveros, Paola.~A. Saavedra, Hector~Javier Hortua, Jorge Garcia~Farieta, and Ivan Olier.
\newblock Conditional diffusion-flow models for generating 3d cosmic density fields: applications to f(r) cosmologies.
\newblock \emph{Machine Learning: Science and Technology}, August 2025.
\newblock ISSN 2632-2153.
\newblock \doi{10.1088/2632-2153/adf8b1}.
\newblock URL \url{http://dx.doi.org/10.1088/2632-2153/adf8b1}.

\bibitem[Rosenblatt(1958)]{rosenblatt1958perceptron}
Frank Rosenblatt.
\newblock The perceptron: a probabilistic model for information storage and organization in the brain.
\newblock \emph{Psychological review}, 65\penalty0 (6):\penalty0 386, 1958.

\bibitem[Saha et~al.(2008)Saha, Prunet, Jain, and Souradeep]{Saha:2007gf}
Rajib Saha, Simon Prunet, Pankaj Jain, and Tarun Souradeep.
\newblock {CMB anisotropy power spectrum using linear combinations of WMAP maps}.
\newblock \emph{Phys. Rev. D}, 78:\penalty0 023003, 2008.
\newblock \doi{10.1103/PhysRevD.78.023003}.

\bibitem[Seljak and Zaldarriaga(1997)]{PhysRevLett.78.2054}
Uros\ifmmode \breve{}\else~\u{}\fi{} Seljak and Matias Zaldarriaga.
\newblock Signature of gravity waves in the polarization of the microwave background.
\newblock \emph{Phys. Rev. Lett.}, 78:\penalty0 2054--2057, Mar 1997.
\newblock \doi{10.1103/PhysRevLett.78.2054}.
\newblock URL \url{https://link.aps.org/doi/10.1103/PhysRevLett.78.2054}.

\bibitem[Smoot(1999)]{smoot1999synchrotronradiationcmbforeground}
George~F. Smoot.
\newblock Synchrotron radiation as cmb foreground, 1999.
\newblock URL \url{https://arxiv.org/abs/astro-ph/9902201}.

\bibitem[Song and Ermon(2020)]{song2020generativemodelingestimatinggradients}
Yang Song and Stefano Ermon.
\newblock Generative modeling by estimating gradients of the data distribution, 2020.
\newblock URL \url{https://arxiv.org/abs/1907.05600}.

\bibitem[Song et~al.(2020)Song, Sohl{-}Dickstein, Kingma, Kumar, Ermon, and Poole]{DBLP:journals/corr/abs-2011-13456}
Yang Song, Jascha Sohl{-}Dickstein, Diederik~P. Kingma, Abhishek Kumar, Stefano Ermon, and Ben Poole.
\newblock Score-based generative modeling through stochastic differential equations.
\newblock \emph{CoRR}, abs/2011.13456, 2020.
\newblock URL \url{https://arxiv.org/abs/2011.13456}.

\bibitem[Song et~al.(2021)Song, Sohl-Dickstein, Kingma, Kumar, Ermon, and Poole]{song2021scorebasedgenerativemodelingstochastic}
Yang Song, Jascha Sohl-Dickstein, Diederik~P. Kingma, Abhishek Kumar, Stefano Ermon, and Ben Poole.
\newblock Score-based generative modeling through stochastic differential equations, 2021.
\newblock URL \url{https://arxiv.org/abs/2011.13456}.

\bibitem[Steier et~al.(2025)Steier, Ghosh, and Delabrouille]{steier2025unbiasedprimordialgravitationalwave}
Alexander Steier, Shamik Ghosh, and Jacques Delabrouille.
\newblock Unbiased primordial gravitational wave inference from the cmb with smica, 2025.
\newblock URL \url{https://arxiv.org/abs/2510.26767}.

\bibitem[Sudevan and Chen(2024)]{Sudevan:2024hwq}
Vipin Sudevan and Pisin Chen.
\newblock {PUREPath: A Deep Latent Variational Model for Estimating CMB Posterior over Large Angular Scales of the Sky}.
\newblock 6 2024.

\bibitem[Sudevan and Saha(2018)]{Sudevan:2017una}
Vipin Sudevan and Rajib Saha.
\newblock {A Global ILC Approach in Pixel Space over Large Angular Scales of the Sky using CMB Covariance Matrix}.
\newblock \emph{Astrophys. J.}, 867\penalty0 (1):\penalty0 74, 2018.
\newblock \doi{10.3847/1538-4357/aae439}.

\bibitem[Thorne et~al.(2017)Thorne, Dunkley, Alonso, and Næss]{Thorne_2017}
B.~Thorne, J.~Dunkley, D.~Alonso, and S.~Næss.
\newblock The python sky model: software for simulating the galactic microwave sky.
\newblock \emph{Monthly Notices of the Royal Astronomical Society}, 469\penalty0 (3):\penalty0 2821–2833, May 2017.
\newblock ISSN 1365-2966.
\newblock \doi{10.1093/mnras/stx949}.
\newblock URL \url{http://dx.doi.org/10.1093/mnras/stx949}.

\bibitem[Vincent(2011)]{6795935}
Pascal Vincent.
\newblock A connection between score matching and denoising autoencoders.
\newblock \emph{Neural Computation}, 23\penalty0 (7):\penalty0 1661--1674, 2011.
\newblock \doi{10.1162/NECO_a_00142}.

\bibitem[Wang(2017)]{Wang2017PrimordialGW}
Mike~S Wang.
\newblock Primordial gravitational waves from cosmic inflation.
\newblock 2017.
\newblock URL \url{https://api.semanticscholar.org/CorpusID:54028419}.

\bibitem[Weiland et~al.(2022)Weiland, Addison, Bennett, Halpern, and Hinshaw]{Weiland_2022}
Janet~L. Weiland, Graeme~E. Addison, Charles~L. Bennett, Mark Halpern, and Gary Hinshaw.
\newblock Polarized synchrotron foreground assessment for cmb experiments.
\newblock \emph{The Astrophysical Journal}, 936\penalty0 (1):\penalty0 24, August 2022.
\newblock ISSN 1538-4357.
\newblock \doi{10.3847/1538-4357/ac83ab}.
\newblock URL \url{http://dx.doi.org/10.3847/1538-4357/ac83ab}.

\bibitem[Yadav and Saha(2021)]{Yadav_2021}
Sarvesh~Kumar Yadav and Rajib Saha.
\newblock A bayesian ilc method for cmb b-mode posterior estimation and reconstruction of primordial gravity wave signal.
\newblock \emph{The Astrophysical Journal}, 914\penalty0 (2):\penalty0 119, June 2021.
\newblock ISSN 1538-4357.
\newblock \doi{10.3847/1538-4357/abfd9b}.
\newblock URL \url{http://dx.doi.org/10.3847/1538-4357/abfd9b}.

\bibitem[Yan et~al.(2023)Yan, Wang, Li, and Xia]{Yan:2023oan}
Ye-Peng Yan, Guo-Jian Wang, Si-Yu Li, and Jun-Qing Xia.
\newblock {Delensing of Cosmic Microwave Background Polarization with Machine Learning}.
\newblock \emph{Astrophys. J. Suppl.}, 267\penalty0 (1):\penalty0 2, 2023.
\newblock \doi{10.3847/1538-4365/acd2ce}.

\bibitem[Yan et~al.(2024)Yan, Li, Wang, Zhang, and Xia]{yan2024cmbfscnncosmicmicrowavebackground}
Ye-Peng Yan, Si-Yu Li, Guo-Jian Wang, Zirui Zhang, and Jun-Qing Xia.
\newblock Cmbfscnn: Cosmic microwave background polarization foreground subtraction with convolutional neural network, 2024.
\newblock URL \url{https://arxiv.org/abs/2406.17685}.

\bibitem[Yan et~al.(2025)Yan, Li, Liu, Xia, and Li]{yan2025foregroundremovalgroundbasedcmb}
Ye-Peng Yan, Si-Yu Li, Yang Liu, Jun-Qing Xia, and Hong Li.
\newblock Foreground removal in ground-based cmb observations using a transformer model, 2025.
\newblock URL \url{https://arxiv.org/abs/2502.09071}.

\bibitem[Ysard et~al.(2010)Ysard, Miville-Deschênes, and Verstraete]{Ysard_2010}
N.~Ysard, M.~A. Miville-Deschênes, and L.~Verstraete.
\newblock Probing the origin of the microwave anomalous foreground.
\newblock \emph{Astronomy and Astrophysics}, 509:\penalty0 L1, January 2010.
\newblock ISSN 1432-0746.
\newblock \doi{10.1051/0004-6361/200912715}.
\newblock URL \url{http://dx.doi.org/10.1051/0004-6361/200912715}.

\bibitem[Zhao et~al.(2023)Zhao, Ting, Diao, and Mao]{zhao2023diffusionmodelconditionallygenerate}
Xiaosheng Zhao, Yuan-Sen Ting, Kangning Diao, and Yi~Mao.
\newblock Can diffusion model conditionally generate astrophysical images?, 2023.
\newblock URL \url{https://arxiv.org/abs/2307.09568}.

\bibitem[Zonca et~al.(2021)Zonca, Thorne, Krachmalnicoff, and Borrill]{Zonca_2021}
Andrea Zonca, Ben Thorne, Nicoletta Krachmalnicoff, and Julian Borrill.
\newblock The python sky model 3 software.
\newblock \emph{Journal of Open Source Software}, 6\penalty0 (67):\penalty0 3783, November 2021.
\newblock ISSN 2475-9066.
\newblock \doi{10.21105/joss.03783}.
\newblock URL \url{http://dx.doi.org/10.21105/joss.03783}.

\end{thebibliography}

\end{document}